\documentclass[12pt,preprint]{aastex}
\usepackage{graphicx,color}
\usepackage{comment}
\shortauthors{Allers \& Liu}
\shorttitle{NIR Spectra of Young Field Dwarfs}

\begin{document}

\newcommand{\Ks}{\mbox{$K_S$}}
\newcommand{\Lp}{\mbox{$L^{\prime}$}}
\newcommand{\mbol}{\mbox{$m_{\rm{bol}}$}}
\newcommand{\degs}{\mbox{$^{\circ}$}}
\newcommand{\perpix}{\mbox{pixel$^{-1}$}}
\newcommand{\mjup}{\mbox{$M_{\rm{Jup}}$}}
\newcommand{\etal}{et al.}
\newcommand{\eg}{e.g.}
\newcommand{\ie}{i.e.}
\newcommand{\htwoo}{{\hbox{H$_2$O}}}   

\def\substitute@option#1#2{%
 \ClassWarning{aastex}{%
  Substyle #1 is deprecated in aastex.
  Using #2 instead (please fix your document).
 }\@nameuse{ds@#2}%
}%
\newcommand\ionscript[2]{#1$\;${\tiny\rmfamily{#2}}\relax}%
\newcommand\ionfootnote[2]{#1$\;${\scriptsize\rmfamily{#2}}\relax}%

\title{A Near-Infrared Spectroscopic Study of \linebreak Young Field Ultracool Dwarfs}

\author{K.~N.~Allers\altaffilmark{1}}
\affil{Department of Physics and Astronomy, Bucknell University, Lewisburg, PA 17837, USA; k.allers@bucknell.edu}

\author{Michael~C.~Liu\altaffilmark{1}} 
\affil{Institute for Astronomy, University of Hawaii, 
2680 Woodlawn Drive, Honolulu, HI 96822, USA}

\altaffiltext{1}{Visiting Astronomer at the Infrared Telescope Facility,
  which is operated by the University of Hawaii under Cooperative
  Agreement No. NCC 5-538 with the National Aeronautics and Space
  Administration, Office of Space Science, Planetary Astronomy Program.}

\begin{abstract}
  We present a near-infrared (0.9--2.4~\micron)
  spectroscopic study of 73 field ultracool dwarfs having spectroscopic
  and/or kinematic evidence of youth ($\approx$10--300~Myr). 
  Our sample is composed of 48~low-resolution ($R \approx$100) spectra
  and 41~moderate-resolution spectra ($R \gtrsim$750--2000).
  First, we establish a method for spectral typing M5--L7 dwarfs at
  near-IR wavelengths that is independent of gravity. We find that both
  visual and index-based classification in the near-IR provide
  consistent spectral types with optical spectral types, though with a
  small systematic offset in the case of visual classification at
  $J$~and $K$~band.
  Second, we examine features in the spectra of $\sim$10~Myr ultracool
  dwarfs to define a set of gravity-sensitive indices 
  based on FeH, VO, \ion{K}{1}, \ion{Na}{1}, and $H$-band continuum
  shape.  
  %
  We then create an index-based method for classifying the gravities of
  M6--L5 dwarfs that provides consistent results with gravity
  classifications from optical spectroscopy.  Our index-based classification can distinguish between young and dusty objects.
  Guided by the resulting classifications, we propose a set of
  low-gravity spectral standards for the near-IR.  Finally, we estimate
  the ages corresponding to our gravity classifications.
\end{abstract}

\keywords{brown dwarfs -infrared:stars - planets and satellites: atmospheres - stars:low-mass}

\section{Introduction}

Brown dwarfs occupy the mass range between stars and planets.  Because they are not massive enough to sustain hydrogen burning in their cores, they continually cool over their lifetimes.  In addition, brown dwarfs contract as they age, evolving from low to high surface gravity.  Thus, brown dwarfs do not occupy a main sequence, and the spectral type of a brown dwarf can not provide a unique determination of its mass.  

The dividing line between brown dwarfs and exoplanets is customarily taken to be the minimum mass at which an object fuses deuterium \citep[11.4--14.4 $M_{\rm{Jupiter}}$;][]{spiegel11}.  This mass boundary does not discriminate between the possible origins of the object.  Is NGC 4349 127 b, a 20~\mjup\ radial velocity companion \citep{lovis07} to an intermediate-mass star (3.9~ $M_{\odot}$), best characterized as a brown dwarf?  Should Cha 110913-773444, an 8~\mjup\ free-floating object with a circumstellar disk \citep{luhman05}, be considered a planet?  The dividing line between planets and brown dwarfs has been further blurred by the discovery of directly imaged exoplanets \citep{marois08, lagrange09}.   In particular, the HR~8799 planets have very red near-IR colors, similar to the reddest known field L-type brown dwarfs.  The remarkably red colors of some L dwarfs have been attributed to youth \citep{kirkpatrick08, barman11} and/or an unusually dusty photosphere \citep{mclean03, cushing05, looper08}.  It is interesting to note that spectroscopy and photometry of the HR~8799 planets are best matched by young, dusty atmospheres \citep[e.g.][]{bowler10,madhusudhan11,barman11}.  
Thus, better understanding of the properties of young and/or dusty field brown dwarfs may provide important insights on the atmospheres of directly imaged exoplanets.

A number of field brown dwarfs having spectroscopic indicators of youth have been reported \citep[e.g.][]{reid08, cruz04}.  To date, these studies have mainly focused on the optical spectroscopic properties of low-gravity M and L field dwarfs \citep{cruz09, kirkpatrick08}.  Only a handful of young field brown dwarfs have been studied in detail in the near-infrared \citep[e.g.][]{allers10, kirkpatrick06}.  There are a number of reasons to study young field brown dwarfs in the infrared.  First, the spectral energy distributions (SEDs) of late-M and L dwarfs peak in the near-infrared, making them easiest to study at these wavelengths.  Second,  studies of directly imaged exoplanets have been conducted in the near-IR, where the planet-to-star flux ratio is most favorable and ground-based adaptive optics systems have the best performance.  Thus, direct spectroscopic and photometric comparison of exoplanets and brown dwarfs is feasible in the near-IR \citep[e.g.][]{bowler10}.  Finally, the near-IR spectra of the youngest brown dwarfs contain a wealth of gravity-sensitive features \citep{allers07}.

In this paper, we present the largest near-IR spectroscopic sample to date of young ($\lesssim$200~Myr old) field ultracool dwarfs.   Our sample consists of 73 late-M and L-type dwarfs displaying spectral signatures of youth and/or kinematic evidence for membership in a young moving group.  At young ages ($\lesssim$125~Myr old), objects with spectral types later than M6.5 are below the hydrogen burning mass limit \citep{stauffer98}.  Thus, the vast majority of our sample is comprised of brown dwarfs.  In this work, we determine near-IR spectral types and examine gravity (age) sensitive features in low and moderate resolution near-IR spectra of young field brown dwarfs.  We develop a set of indices to classify the gravity of these objects and create a gravity classification system for use in the near-IR that provides results consistent with gravity classifications from optical spectroscopy \citep{cruz09}.  We then compare our findings on young field objects to objects with well-determined ages.

\section{Our Sample}

Our sample consists of 89 spectra for 73 M5--L7 objects that have published spectroscopic or kinematic evidence of youth (see Table~\ref{tbl:spt} and references therein).  Thirty of our spectra are for objects having gravity determinations from optical spectroscopy.  \citet{cruz09} propose optical gravity classifications of $\delta$, $\gamma$, and $\beta$ and suggest that they correspond to ages of $\sim$1, $\sim$10, and $\sim$100~Myr, respectively.  Included in our sample are nine spectra for members of the $\sim$12~Myr old TW~Hydra moving group (hereinafter TWA).  Six spectra in our sample are substellar companions to young stars.  

In addition to our sample of young objects, we also use a sample of field dwarfs having no known spectral peculiarities to establish a normal gravity sequence.  At low-spectral resolution ($R \approx$100), we include the spectra of field dwarf standards from \citet{geissler11} and \citet{kirkpatrick10} as well as the spectral templates from \citet{burgasser10}.  At moderate resolution ($R \approx$750--2000), we use spectra from the IRTF Spectral Library \citep{cushing05}.

We also include published spectra of objects thought to have dusty photospheres \citep{kirkpatrick10, looper08}.  These objects have peculiar near-IR spectra and very red near-IR colors but are thought to have normal gravity based on their optical spectra.  These dusty objects provide an important test for our indices and classification system.  Our gravity-sensitive indicators should show these dusty objects to have normal gravities, similar to field dwarfs.

\section{IRTF/SpeX Near-IR Spectroscopy}

\begin{figure}
\vskip 0.5in
\centerline{\includegraphics{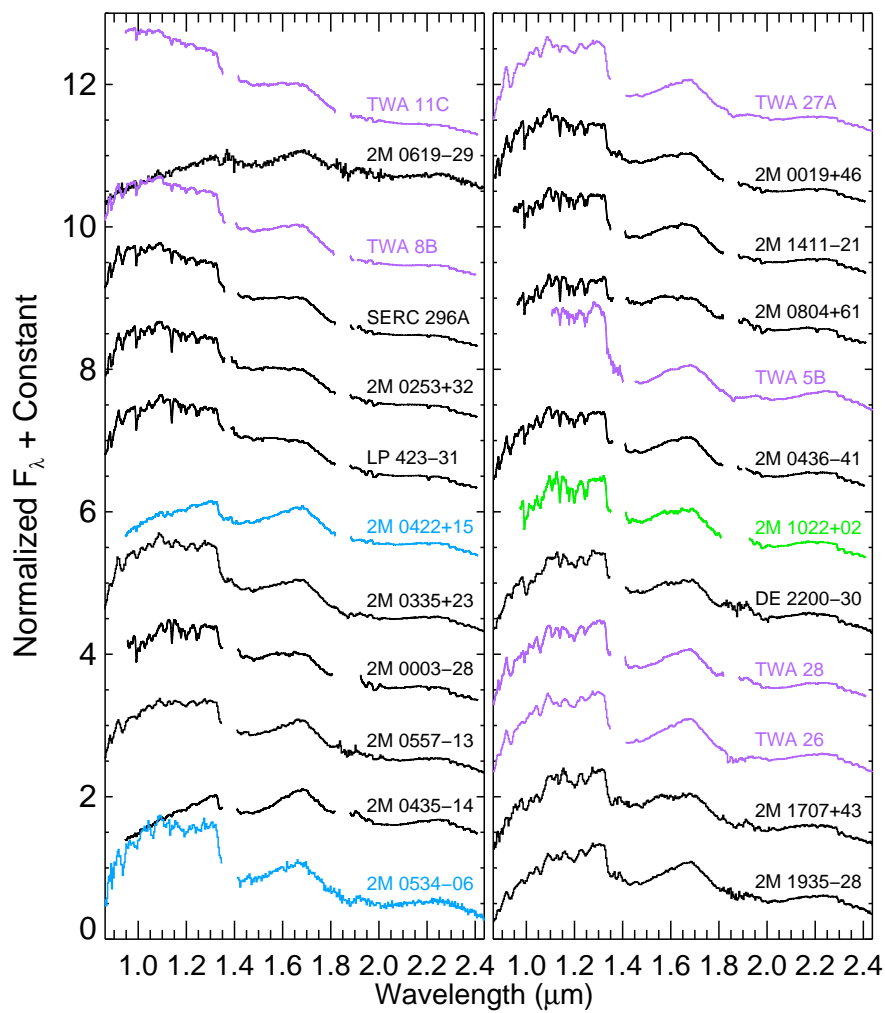}}
\vskip 2ex
\caption[Near-IR Spectra of a Large Sample of Young Field Dwarfs: M Dwarfs]
{\label{mallspec} Near-IR spectra of the M dwarfs in our sample.  Moderate-resolution spectra have been smoothed with a Gaussian to a resolution of $\sim$200 for display purposes.  For objects having both low- and moderate-resolution spectra, the low-resolution spectrum is displayed.  Spectra plotted in purple are known members of the $\sim$12~Myr old TW Hydra Association (TWA).  Spectra plotted in blue have optical gravity classifications of $\gamma$ or $\delta$.  Spectra plotted in green have optical gravity classifications of $\beta$.  Objects plotted in black have no available optical gravity classification.}
\end{figure}

\begin{figure}
\vskip 0.5in
\rotate
\centerline{\includegraphics[angle=90, origin=c]{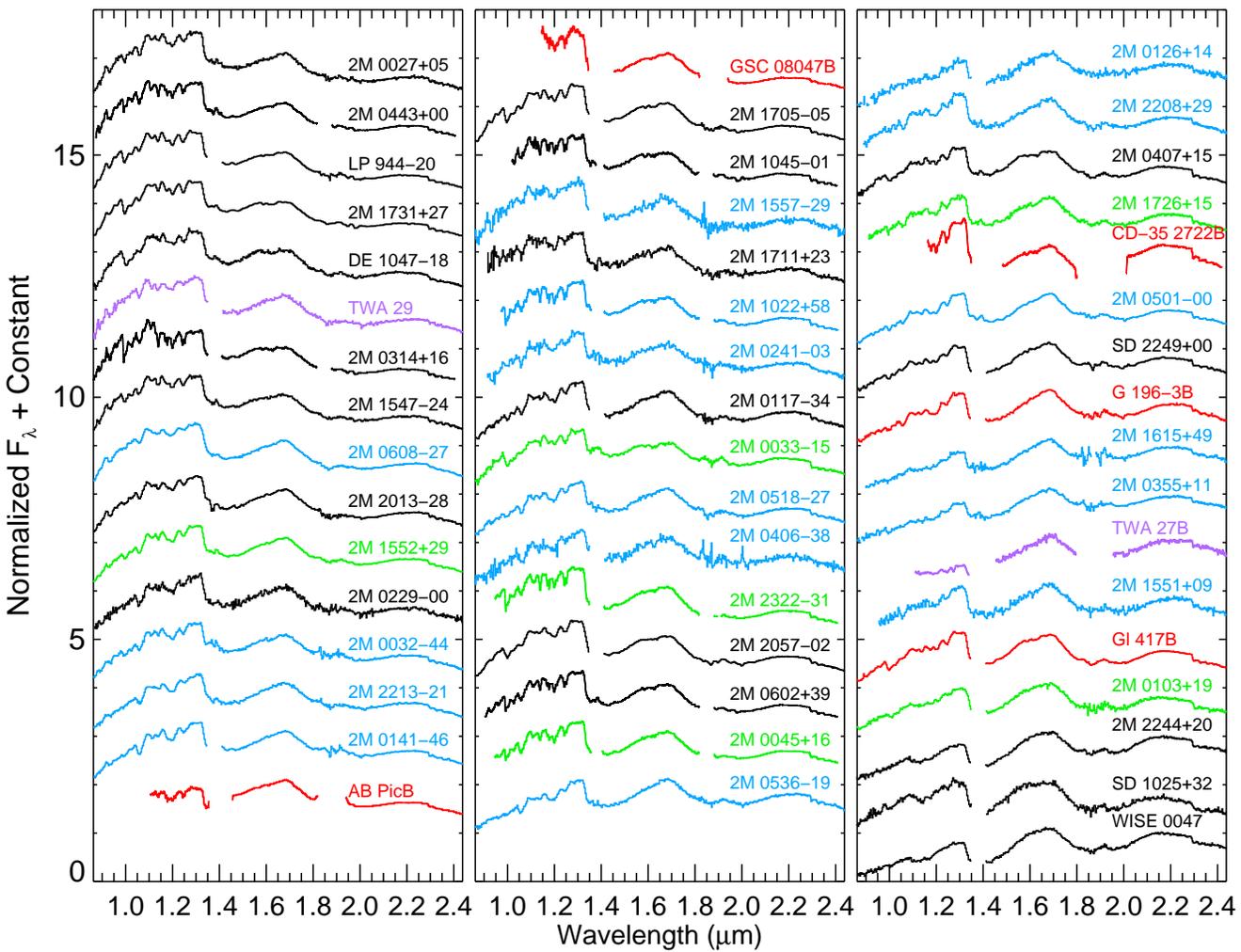}}
\vskip 2ex
\caption[Near-IR Spectra of a Large Sample of Young Field Dwarfs: L Dwarfs]
{\label{lallspec} Near-IR spectra of the L dwarfs in our sample.  Color coding is the same as for Figure \ref{mallspec}, and young companion spectra are displayed in red.}
\end{figure}

We obtained spectroscopy of our targets using the SpeX spectrograph \citep{rayner03} on the 3~m  NASA Infrared Telescope Facility (hereinafter IRTF) located on the summit of Mauna Kea, Hawai'i.  Our sample includes observations taken using the 0.8--2.4~$\mu$m moderate-resolution ($R \approx$750--2000), cross-dispersed mode (hereinafter SXD) and the 0.8--2.5~$\mu$m low-resolution ($R \approx$100), prism mode (hereinafter PRZ).  The instrument configurations, integration times and observation dates for each of our targets are listed in Table~\ref{tbl:spex}.  Data reduction was performed using the facility reduction pipeline, Spextool \citep{cushing04}.  We observed A0 stars proximate in time and sky position to our science targets and used these data to correct for telluric absorption following the method described in \citet{vacca03}.  We also added published near-IR spectra to our sample (references listed in Table~\ref{tbl:spt}).  Spectra of our sample are displayed in Figures~\ref{mallspec} and \ref{lallspec}.

\section{Analysis}

\subsection{Spectral Typing}

Spectral classification is primarily based on visual examination of a
large sample of objects in a common wavelength range and then choosing
specific objects to serve as the defining templates for the spectral
classes and subclasses. This is the heart of the long-established
Morgan--Keenan system, spanning on the two parameters of overall spectral
appearance (\ie, spectral type) and luminosity class \citep{morgan43}. 
A visual-based approach has several advantages, most
notably that the ensemble morphological information of the data is best
captured by human judgement. Practically speaking, a well-constructed
sequence should result in a smooth progression of changes in spectral
morphology \citep[e.g., see discussion in][]{kirkpatrick05}.

For field late-M and L dwarfs, there exist well-established spectral
classification systems for optical (far-red) spectra \citep{boeshaar85,kirkpatrick99}. 
At these wavelengths and spectral
types, multiple absorption features are gravity-sensitive and have been
used to identify young objects in young clusters and the field \citep[\eg, ][]{martin99a, slesnick04, kirkpatrick06}. \citet{cruz09} 
have proposed a formal system for classifying young L0--L5
dwarfs at optical wavelengths based on template objects spanning three
gravity classes ($\alpha$, $\beta$, and $\gamma$). Optical spectral
types for our sample are listed in Table~\ref{tbl:spt} and typically
have an uncertainty of 1 subtype.

However, young ultracool dwarfs lack any classification system at the
near-IR wavelengths, both for spectral type and gravity. Development of
such a system is compelling for two reasons. Late-M and L~dwarfs are
brightest in the near-IR, making spectroscopic followup possible with
moderate-aperture (3--4~m) telescopes. Moreover, previous studies have
shown that the appearance of the near-IR continuum is strongly
gravity-dependent, even at low ($R\sim100$) spectral resolution \citep[\eg,][]{lucas01, allers07, lodieu08}. 

The major obstacle to a near-IR system is the heterogeneous nature of
ultracool dwarf spectra at these wavelengths. Unlike the smooth sequence
of optical spectra, the progression of near-IR spectra for L dwarfs is
far more irregular, especially at $H$ and $K$-band \citep[e.g., ][]{kirkpatrick05}.
At fixed optical spectral type, there is also a large spread in the
near-IR colors, believed to be due to the influence of gravity,
metallicity, and photospheric condensate variations \citep[\eg,][]{knapp04, stephens09}. Therefore the overall shapes of the near-IR
SEDs do not follow a smooth sequence, inhibiting the traditional
approach of defining actual template objects.

Our large collection of spectra provides a unique opportunity to examine
classification of ultracool dwarfs in the near-IR. We follow a
two-pronged approach. 

\begin{enumerate}

\item First, we apply visual classification methods by comparing our
  sample with the set of $J$-band spectral standards proposed by
  \citet{kirkpatrick10}. They have found classification is possible
  for field objects in this restricted wavelength regime, with the
  resulting near-IR types being very similar to optical types for the
  same objects. {\em We show here that qualitative (visual-based)
    classification also works for young L~dwarfs, though there is a
    $\approx$1~subclass systematic shift between the near-IR and optical
    types.}

\item Then, we consider a quantitative (index-based) approach to
  classifying spectral types and gravities. Previous near-IR studies
  have shown that the \htwoo\ absorption bands, as measured by flux
  ratios (indices), are well-correlated with optical spectral type
  \citep[\eg, ][]{reid01, geballe02, mclean03}.
  While this approach is philosophically different than visual
  classification, these studies demonstrate that prominent near-IR
  features do change in the same fashion as the overall optical
  spectrum, \ie, they track similar physical changes. (In addition, even
  visually-based systems often rely on indices as a practical recipe for
  classification, \eg, the approach of \citet{kirkpatrick00}
  for L~dwarfs.) So while the overall issues with near-IR classification
  are not solved, it is possible to access information about the
  underlying physical properties of L~dwarfs. Similar to previous
  studies for old field objects, {\em we show that well-chosen indices
    in the near-IR successfully correlate with the spectral type and
    gravity designations in the optical, allowing us to classify near-IR
    spectra in a practical and useful fashion.}

\end{enumerate}

\subsubsection{Visual Classification}

\begin{figure}
\vskip 0.5in
\centerline{\includegraphics[width=16cm]{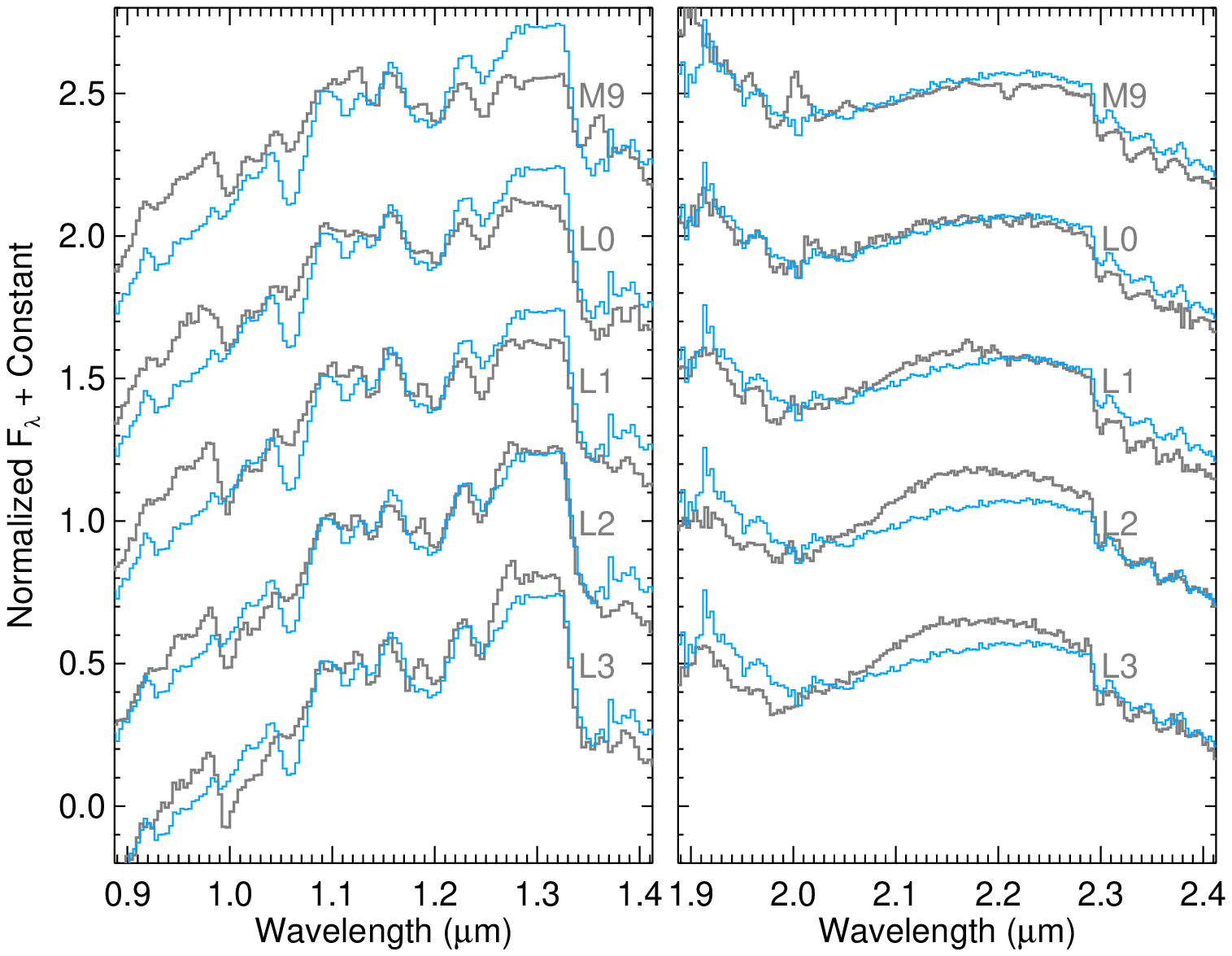}}
\vskip 2ex
\caption[An Example of $J$ and $K$ band Visual Classifications]
{\label{vis_class} An example of visual classification using the spectrum of 2M~0141-46, which is optically classified as L0$\gamma$ \citep{cruz09}. The spectrum of 2M~0141-46 \citep{kirkpatrick06} is plotted in blue.  Spectra of field dwarf standards \citep{kirkpatrick10} are plotted in gray.  Based on visual comparison of the 1.07--1.40~$\mu$m spectrum of 2M~0141-46 to field dwarf standards, we assign a $J$-band spectral type of L2.  The $K$-band spectral type of 2M~0141-46, L0, is assigned based on the best-matching field standard at wavelengths of 1.90--2.20~$\mu$m.}
\end{figure}

We first compared our spectra of young objects with field dwarf near-IR spectral standards from \citet{kirkpatrick10}.  Not surprisingly, the entire 0.8--2.5~$\mu$m spectra of young objects are not well-matched by older field dwarfs, in part due to the redder near-IR colors of young objects \citep{kirkpatrick08}.  In some spectral regions, however, the continuum shape is sensitive to spectral type with little dependence on gravity.  In particular, we used the 1.07--1.40~$\mu$m and 1.90--2.20~$\mu$m wavelength regions to determine $J$-band and $K$-band spectral types by visual comparison to field dwarf standards (Figure \ref{vis_class}).  For each spectrum in our sample, we over-plotted the spectra of field dwarf standards (normalized over the comparison wavelength region) and qualitatively determined which standard best matched the continuum shape of our object.  If two standard spectra provided similar matches, we assigned the spectral type intermediate to the two standards, \eg, if the L0 and L1 standards provided equally good fits, we assigned a spectral type of L0.5.  For the majority of our objects, selecting a standard with 1 subtype difference compared to the best fitting standard provided a noticeably poorer fit.  Thus, we assign an uncertainty of 1 subtype to our visual classifications.  For objects which had either particularly noisy or peculiar spectra, several field dwarf standards provided equally good fits, thus uncertainties of $\pm$2 subtypes were assigned to their visual classification.

\begin{figure}
\vskip 0.5in
\centerline{\includegraphics[width=16cm]{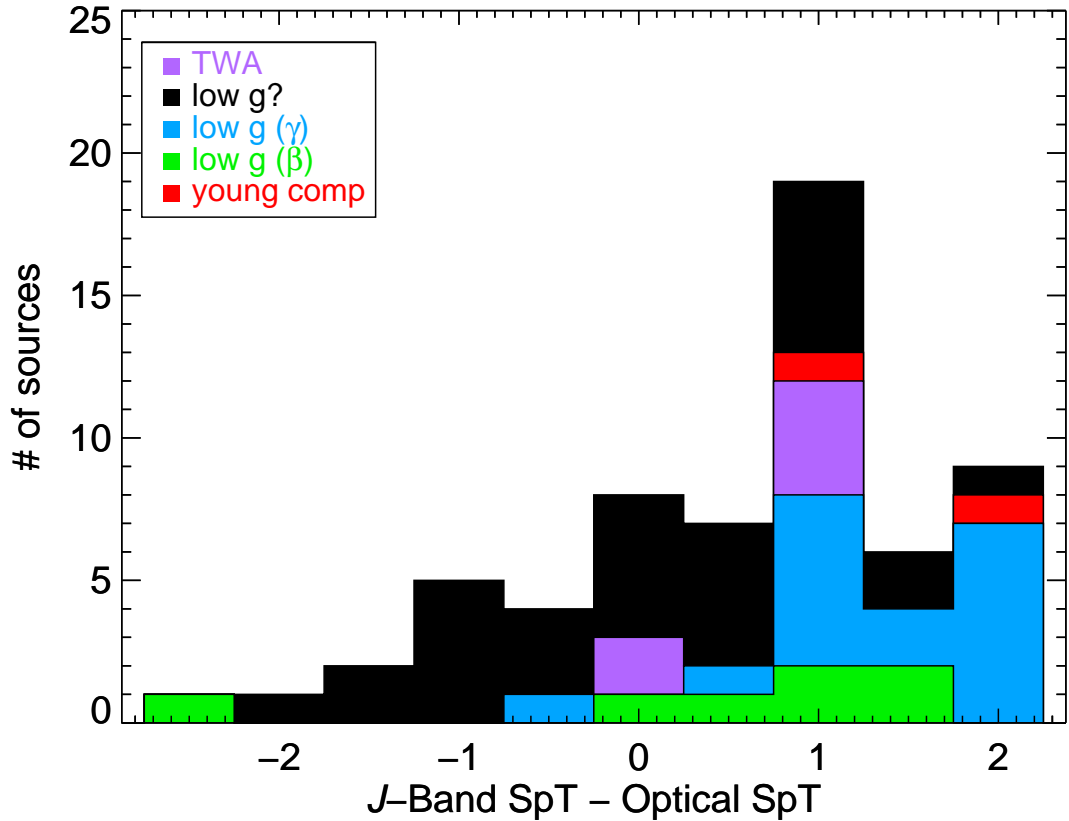}}
\vskip 2ex
\caption[Comparison of Optical and $J$-band Visual Spectral Types]
{\label{jspt_hist} Histogram showing the difference between our $J$-band visual spectral types and published spectral types.  The $J$-band visual spectral types for young objects tend to be later than their optical spectral types, particularly among the lowest gravity objects in our sample (optical classifications of $\gamma$ and TWA members). }
\end{figure}

\begin{figure}
\vskip 0.5in
\centerline{\includegraphics[width=16cm]{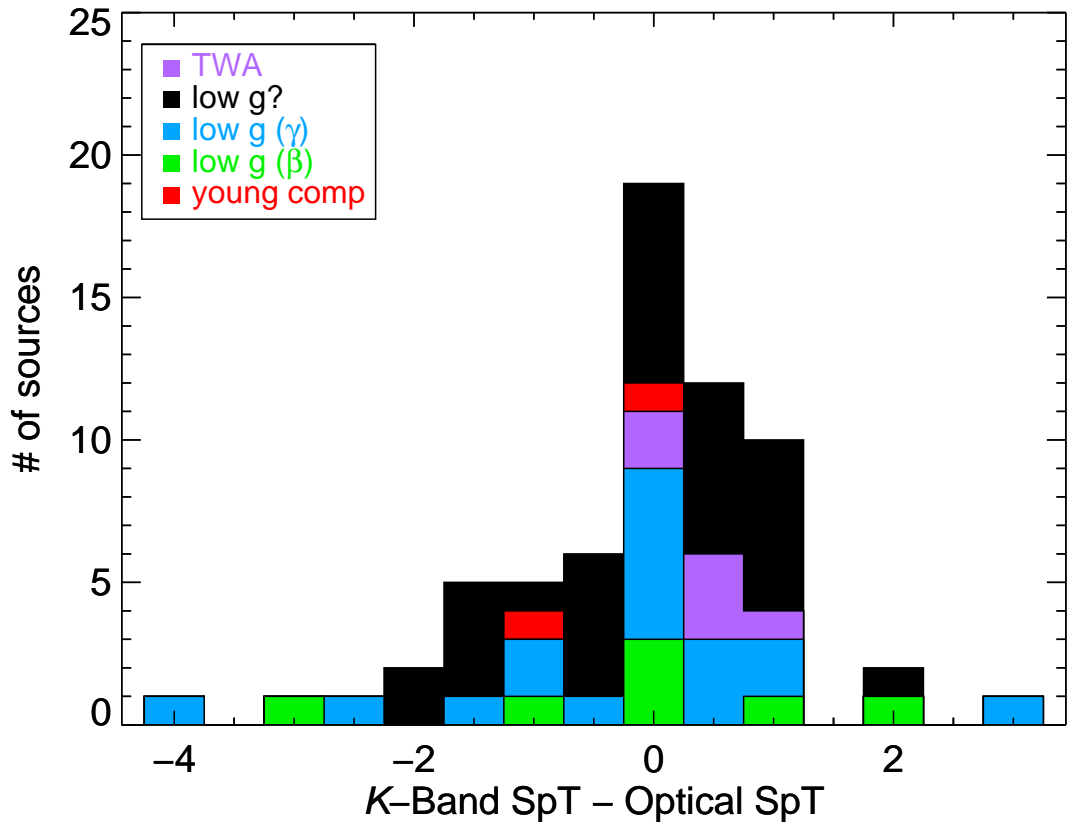}}
\vskip 2ex
\caption[Comparison of Optical and $K$-band Visual Spectral Types]
{\label{kspt_hist} Histogram showing the difference between our $K$-band visual spectral types and published optical spectral types.  For the majority of our sample, our $K$-band spectral types agree with optical spectral types to within our uncertainty ($\pm$1 subtype). }
\end{figure}

To test for any gravity dependence of our visual spectral typing, we compared the near-IR spectral types of $\sim$10 Myr old objects in our sample (TWA members and objects with optical gravity classifications of $\gamma$) to their optical spectral types.  On average, the $J$-band spectral types of $\sim$10 Myr old objects are 1.3 subtypes later than their corresponding optical spectral types.   The $K$-band spectral types for low-gravity objects are, on average, 0.1 subtypes earlier than their corresponding optical spectral types.  \emph{Determining near-IR spectral types for young objects based on visual comparison to field dwarfs will lead to near-IR types that are slightly later than their optical spectral types.}
In addition to visually comparing our spectra to the IR standards, we also computed reduced $\chi ^2$ for the 1.07--1.40~$\mu$m and 1.90--2.20~$\mu$m  
wavelength regions to assign spectral types based on the best matching standard. 
The spectral types determined from the minimum $\chi ^2$ value agreed to within the uncertainties with the spectral types determined by visual classification ($\pm$1 subtype).  Minimum reduced $\chi ^2$ values for fitting spectra of field standards to $\sim$10~Myr old objects were typically $\sim$30 for $J$-band fits and $\sim$7 for $K$-band fits.  The large values for reduced $\chi^2$ are a result of the poor match between field dwarf standards and young field objects.  Figures \ref{jspt_hist} and \ref{kspt_hist} show the differences between our $J$ and $K$-band visual spectral types and published optical spectral types.

\subsubsection{Index-based Classification}

\begin{figure}
\vskip 0.5in
\hskip -1.5in
\centerline{\includegraphics[width=3in]{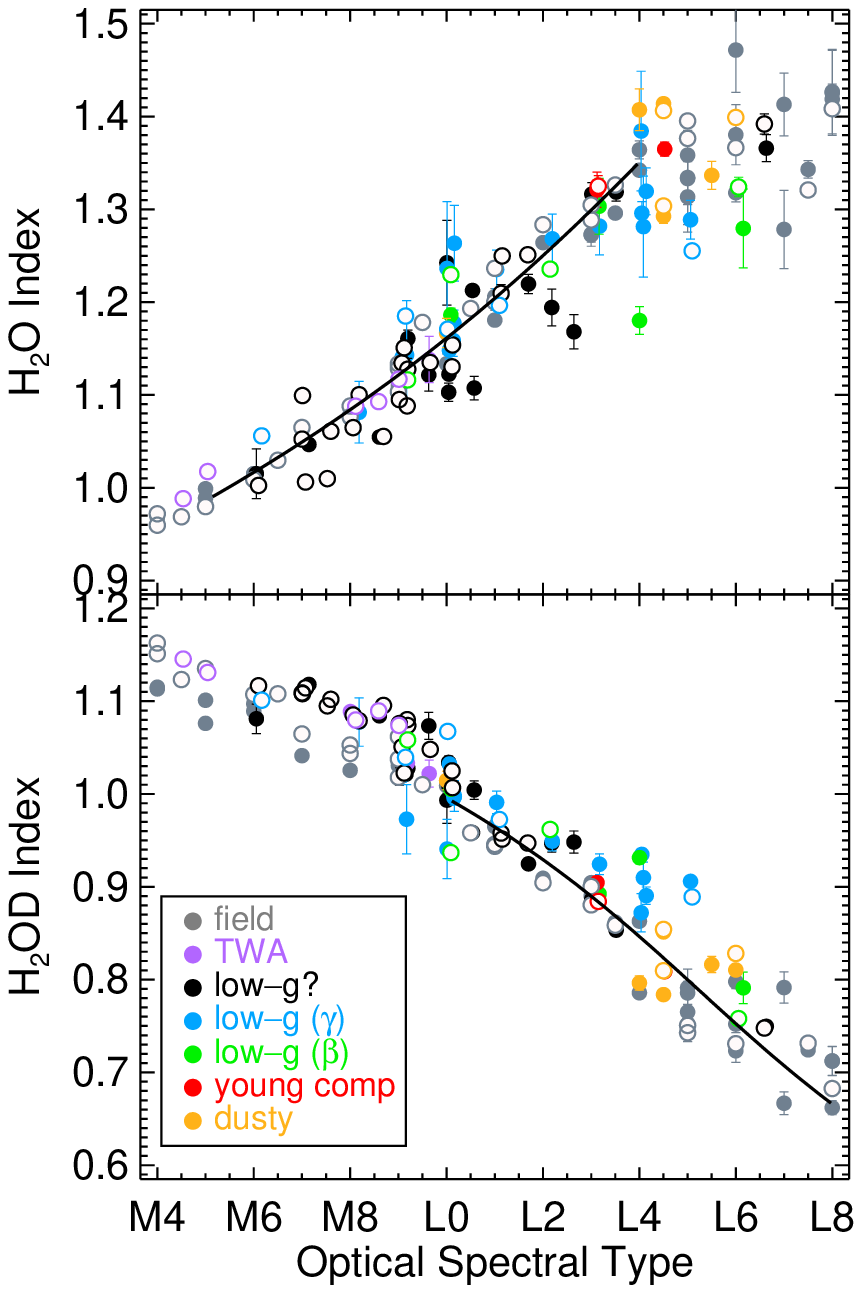}}
\vskip -4.5 in
\hskip +1.5in
\centerline{\includegraphics[width=3in]{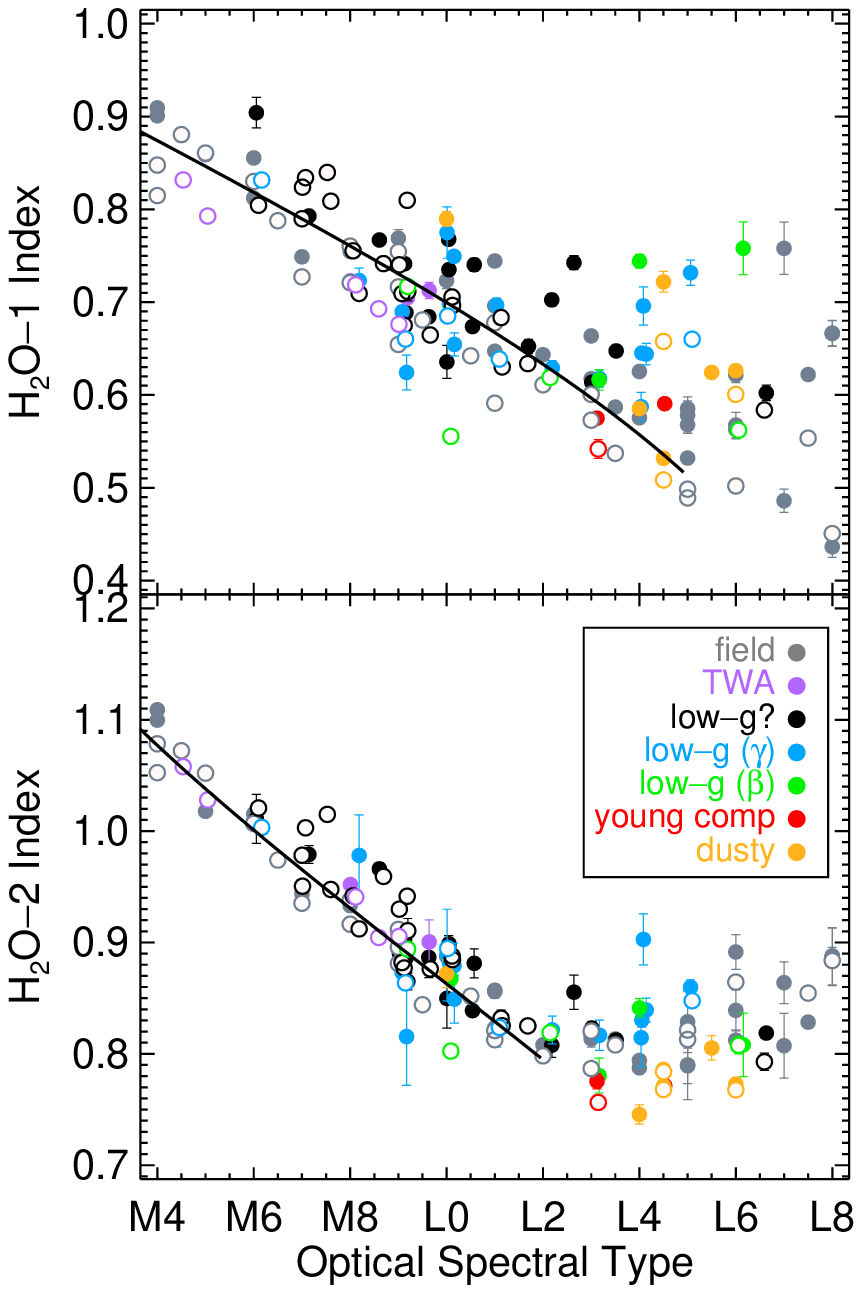}}
\vskip 2ex
\caption{\label{spt} Spectral type sensitive indices as a function of optical spectral type for four near-IR indices showing negligible gravity dependence.  Data and references are presented in Table \ref{tbl:spt}.  Normal field dwarfs are plotted as gray points.  Purple points represent members of the TW~Hydra moving group ($\sim$10 Myr old).  Objects in our sample with an optical gravity classification of $\beta$ are displayed as green points and those having an optical classification of $\gamma$ are displayed as blue points.  Black points (low g?) show objects in our sample having no optical gravity classification.  Red points represent young companions to stars.  Objects with normal gravity but thought to have dusty photospheres are displayed in orange.  Filled circles show index values calculated from low-resolution ($R \approx$100) spectra, and open circles show values calculated from moderate-resolution ($R \approx$750--2000) spectra.  The solid black lines are third-degree polynomial fits to index vs. optical spectral type for field dwarfs and are plotted only over the range of spectral type sensitivity for each index.  Index definitions, ranges and polynomial fits are presented in Table \ref{tbl:indexfits}.}
\end{figure}

Another method of determining spectral type is using spectral indices.  This method has the potential advantage of measuring spectral features which are well-correlated with the overall spectral morphology (type) of the object, while avoiding wavelength regions containing broad, gravity-sensitive features.  We calculated many published spectral type-sensitive indices \citep{tokunaga99, cushing00, testi01, geballe02, mclean03, slesnick04, allers07, weights09, covey10, scholz12} for the objects in our sample.  Most of these spectral-type sensitive indices were developed to correlate with optical spectral types.  The majority of spectral type-sensitive indices were either found to be gravity sensitive (i.e., young dwarfs and field dwarfs having very discrepant index--SpT relations) or were only sensitive over a narrow range in spectral type.  Overall, we find that the H$_2$O \citep{allers07}, H$_2$O-1, H$_2$O-2 \citep{slesnick04} and H$_2$OD\footnote{The H$_2$OD index \citep{mclean03} uses wavelength windows that are smaller than a resolution element of our low resolution spectra, so we widened the index windows to 0.013~$\mu$m (Table~\ref{tbl:indexfits}).} \citep{mclean03} indices are spectral type sensitive and gravity-insensitive over a broad range in spectral type.
For these four indices, we fit index versus optical spectral type for field dwarfs with a third-degree polynomial and use the scatter about the fit as the uncertainty in the index-SpT relation (Figure~\ref{spt}).  The polynomial fits and scatter in the index--SpT relations, along with the range of spectral type sensitivity are presented for each index in Table~\ref{tbl:indexfits}.  Table~\ref{tbl:spt} presents the spectral types calculated from these indices for our sample of young objects.  Uncertainties in the index-derived spectral types were determined using a Monte Carlo approach to account for uncertainties in the index calculations from each spectrum.  These were then added in quadrature to the rms SpT scatter in the index--SpT relations (Table~\ref{tbl:indexfits}).

\subsubsection{Final Near-IR Spectral Types}
\begin{figure}
\vskip 0.5in
\centerline{\includegraphics[width=16cm]{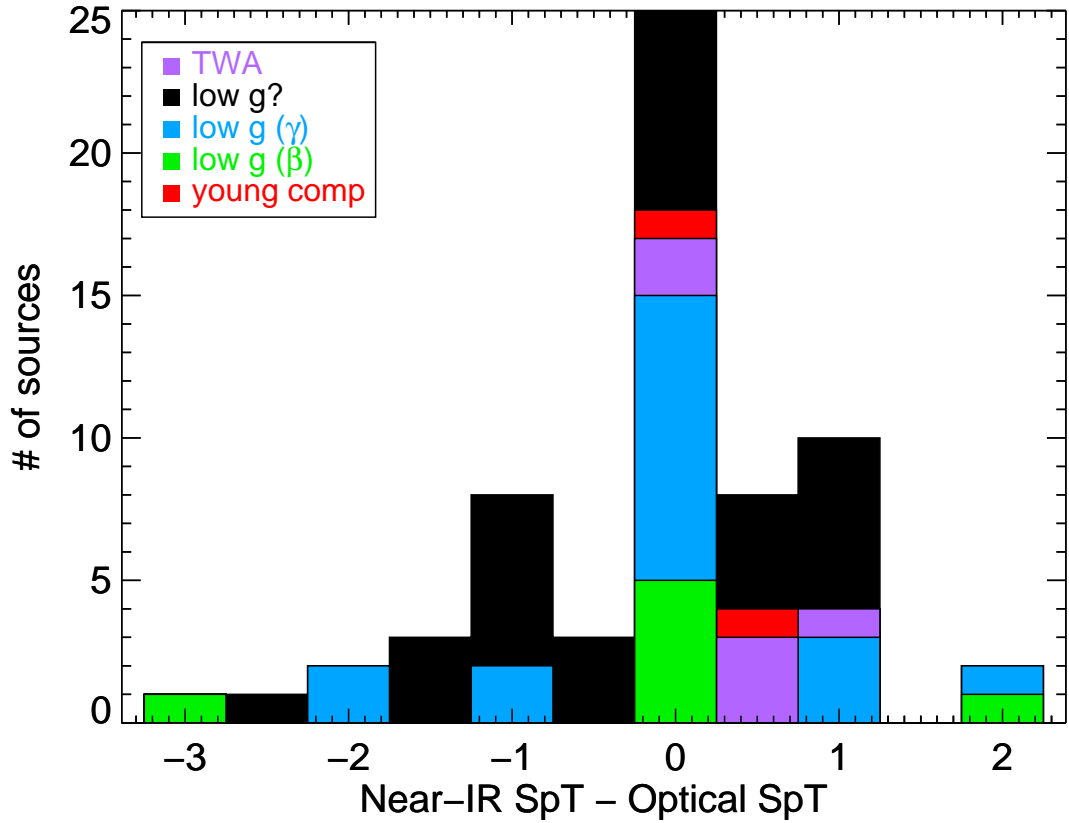}}
\vskip 2ex
\caption[Comparison of Optical and IR Spectral Types]
{\label{spt_hist} Histogram showing the difference between our near-IR spectral types and published optical spectral types.  For the majority of our sample (55 out of 64 objects having optical spectral types), our near-IR spectral types agree with optical spectral types to within our uncertainty ($\pm$1 subtype). }
\end{figure}

\begin{figure}
\vskip 0.5in
\centerline{\includegraphics[width=16cm]{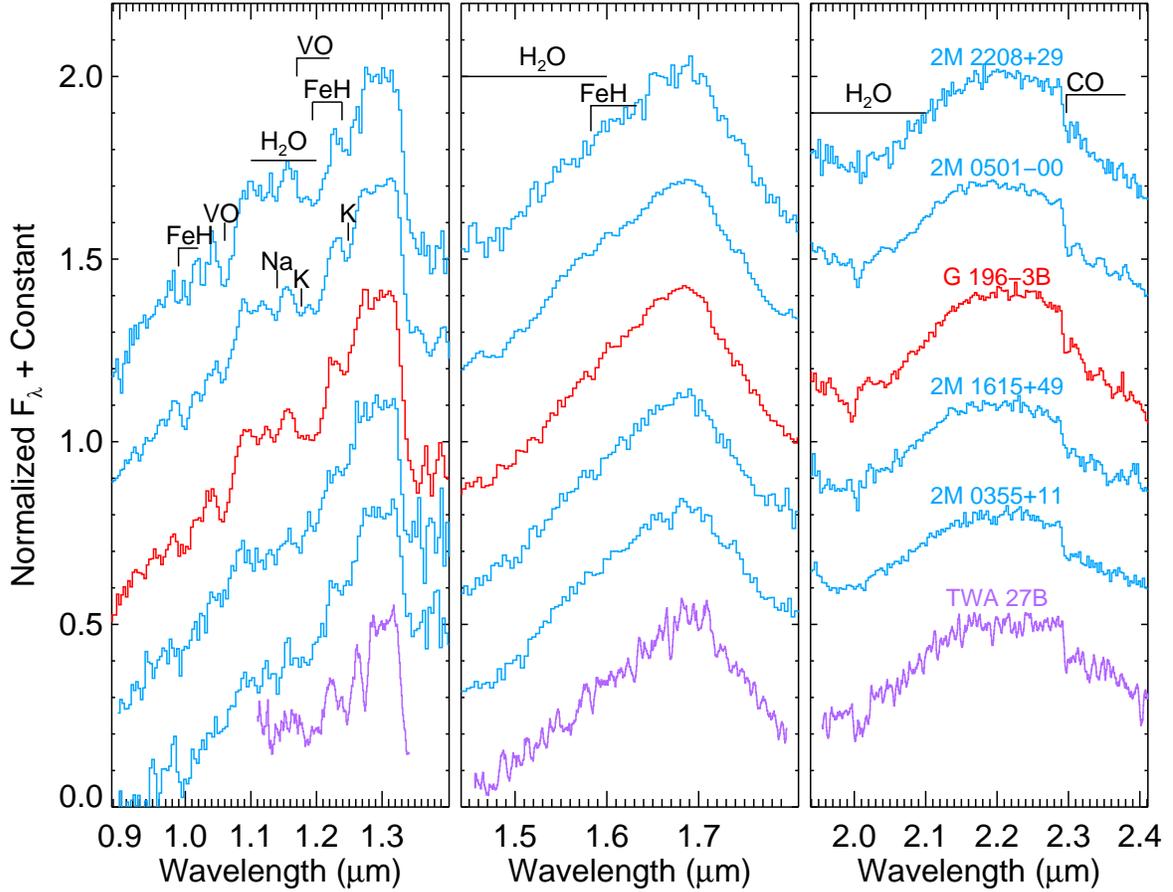}}
\vskip 2ex
\caption[Spectra of very low-gravity L3 dwarfs]
{\label{l3_vlg} Spectra of objects in our sample having a near-IR spectral type of L3 and classified as {\sc vl-g} (Section 4.3).  Spectra plotted in blue have optical gravity classifications of $\gamma$.   The red spectrum is G~196-3B,  a low-mass companion to a young star.  The purple spectrum is TWA 27B (2M 1207b).  The $J$, $H$, and $K$-bands are plotted separately and normalized by the mean flux at 1.27--1.32, 1.65--1.72, and 2.15--2.25~$\mu$m, respectively.  The regions of the spectra where indices are used to measure spectral type (1.29--1.35~$\mu$m, 1.49--1.56~$\mu$m and 1.95--2.09~$\mu$m) are very similar, supporting the L3 IR spectral type assigned to all of them.  Despite the similarity of these spectra in the $H$ and $K$-bands, the $J$-band spectra are more diverse.  In particular, 2M~0355+11 and 2M~1615+49 do not have deep 1.06~$\mu$m VO features.  The spectral shape from $\sim$1.07--1.2~$\mu$m also shows noticeable variation.  The $J-K$ colors for these objects vary from 1.6 (top spectrum) to 3.1 mag (bottom spectrum).}
\end{figure}

To arrive at final near-IR spectral types, we take the weighted mean of all of the spectral types determined using indices and visual comparison.  For many of our objects, this results in the visual spectral types having little effect on the final spectral type determination.  We round the final spectral type to the nearest integer subtype.  Uncertainties in the weighted mean spectral type were 0.3--0.9 subtypes, thus we adopt a conservative uncertainty of 1 subtype for our near-IR spectral types.  The near-IR spectral types of our sample are, on average, 0.07 subtypes earlier than their published optical spectral types.  For the $\sim$10~Myr old objects in our sample, their near-IR spectral types are, on average, identical to their published optical spectral types.  Figure \ref{spt_hist} shows the differences between our near-IR spectral types with published optical spectral types.  The vast majority (55/64) of near-IR and optical spectral types agree to within 1 subtype.  The nine objects having discrepant (by more than 1 subtype) near-IR and optical spectral types do not show a preference for particular optical spectral types, gravities or near-IR colors.  Overall, our method for determining IR spectral types yields results that are consistent with optical spectral types.  An additional benefit of our method of spectral typing is that we are not biased by an object's $J-K$ color, which has a large dispersion among the L dwarfs \citep[\eg, ][]{knapp04}, inhibiting the use of the entire spectrum for visual classification.   Figure \ref{l3_vlg} shows the spectra of objects classified as L3 that have $J-K$ colors ranging from 1.6 to 3.1~mag.  


\subsection{Surface Gravity Indicators}

\begin{figure}
\vskip 0.5in
\centerline{\includegraphics[width=16cm]{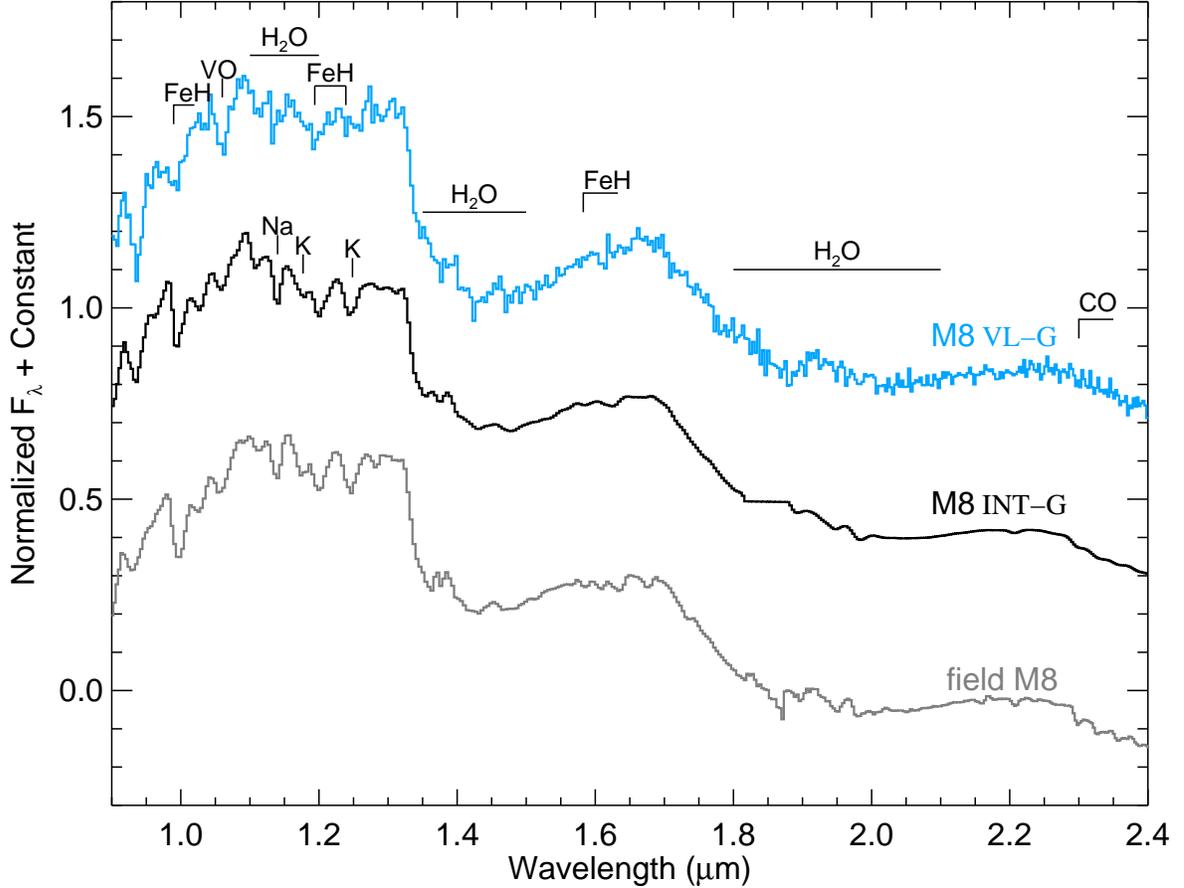}}
\vskip 2ex
\caption[Low-resolution spectra of M8 objects]
{\label{m8_prz} Low-resolution spectra comparing the gravity (age) sensitive features for objects classified as M8 in the near-IR.  Details on the near-IR gravity classifications are described in Section~4.3.  From its optical spectrum, the {\sc vl-g} object (blue; 2M~0534-06) is classified as M8$\gamma$ \citep{kirkpatrick10}.  The {\sc int-g} spectrum is 2M~0019+46 \citep[optical SpT of M8;][]{cruz03} which has been smoothed to a resolution of $\sim$100 to match the low-resolution comparison spectra.  The field dwarf spectrum (gray) is vB~10 \citep{burgasser04}.  The $H$-band continuum shows a distinct triangular shape at low gravities.}
\end{figure}

\begin{figure}
\vskip 0.5in
\centerline{\includegraphics[width=16cm]{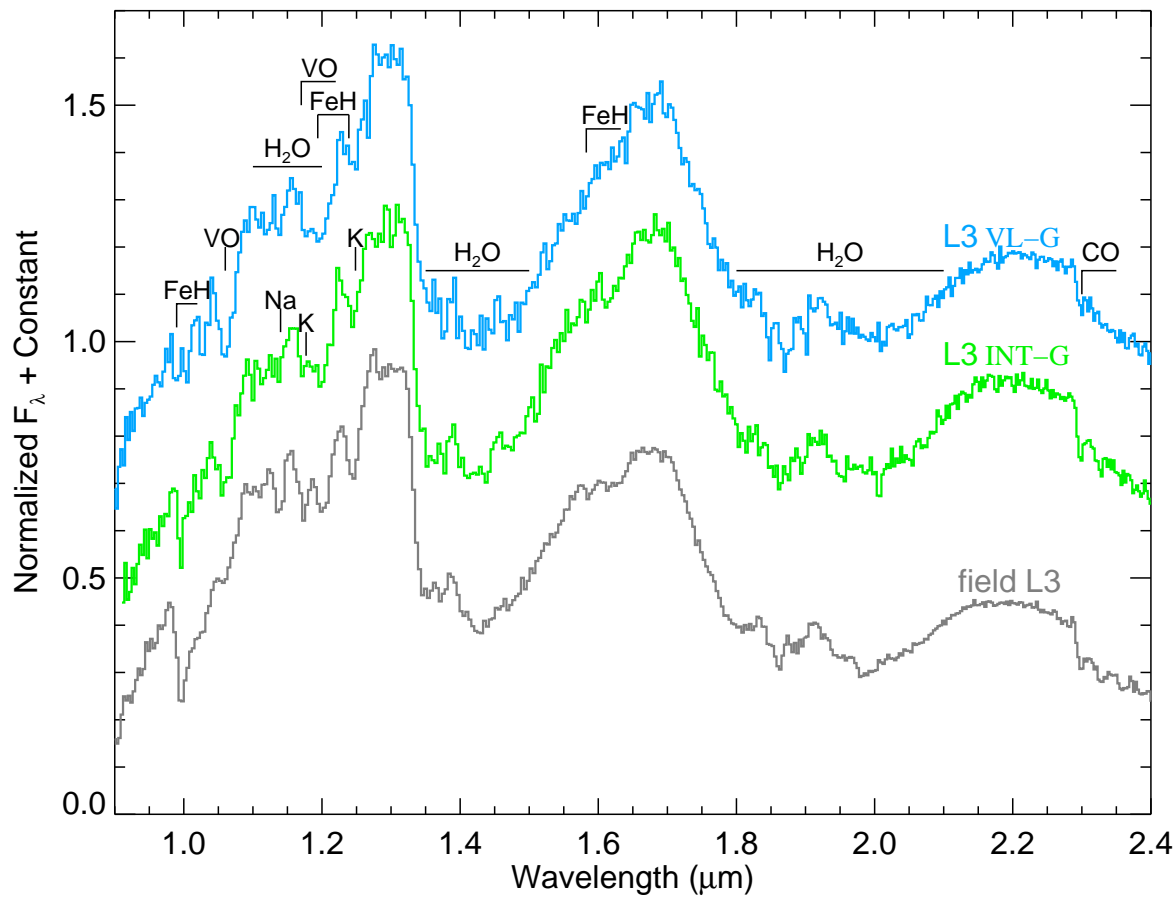}}
\vskip 2ex
\caption[Low-resolution spectra of L3 objects]
{\label{l3_prz} Low-resolution spectra comparing the gravity (age) sensitive features for objects classified as L3 in the near-IR.  From their optical spectra, the {\sc vl-g} object (blue; 2M~2208+29) is classified as L3$\gamma$ and the {\sc int-g} object (green; 2M~1726+15) is classified as L3$\beta$ \citep{cruz09}.  The field dwarf (gray) is the L3 standard 2M~1506+13 \citep{burgasser07c}.  FeH, \ion{Na}{1}, \ion{K}{1}, features are weaker at lower gravities and VO is stronger.  The $H$-band continuum shows a distinct triangular shape at low gravities.}
\end{figure}

\begin{figure}
\vskip 0.5in
\centerline{\includegraphics[width=16cm]{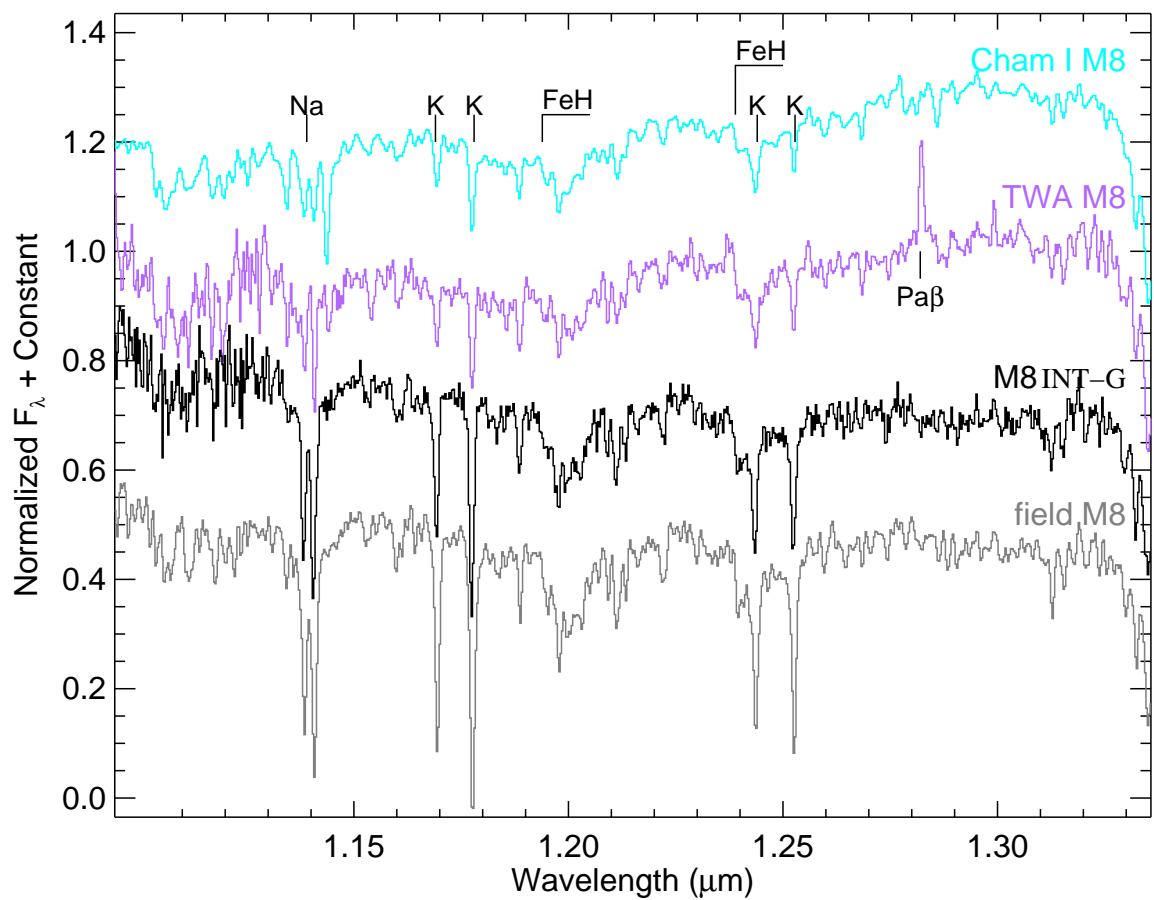}}
\vskip 2ex
\caption[Moderate-resolution spectra of M8 objects]
{\label{m8_sxd} Moderate-resolution spectra comparing the gravity (age) sensitive features in the $J$-band spectra of objects classified as M8 in the near-IR.  The TWA M8 spectrum (purple) is TWA~27, which is classified as {\sc vl-g}.  The gravity classification of the {\sc int-g} spectrum (2M~0019+46) is described in Section 4.3.  Though not an object in our sample, the spectrum of a $\sim$3~Myr old Chamaeleon I M8 (CHSM 17173; K.~L.~Luhman 2007, private communication) is displayed for comparison.  A normal-gravity,  field M8 is also displayed \citep[vB~10;][]{cushing05}.  Young, low-gravity M8s have weaker \ion{Na}{1}, \ion{K}{1} and FeH features than normal field M8 dwarfs.}
\end{figure}

\begin{figure}
\vskip 0.5in
\centerline{\includegraphics[width=12cm]{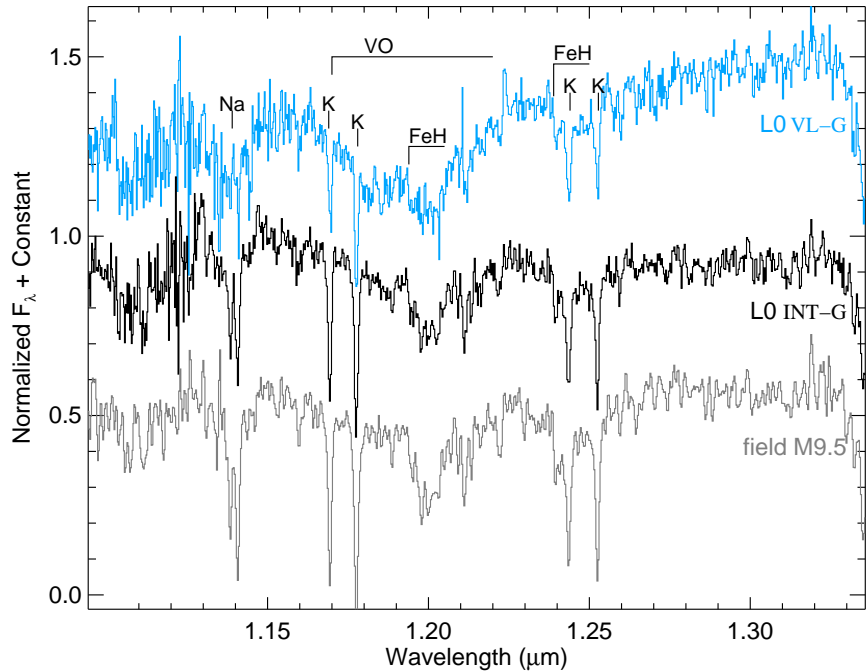}}
\vskip 2ex
\caption[Moderate-resolution spectra of L0 objects]
{\label{l0_sxd} Moderate-resolution spectra comparing the gravity (age) sensitive features in the $J$-band spectra of objects classified as L0 in the near-IR.  The L0 {\sc vl-g} spectrum is 2M~0141-46, which has an optical gravity classification of $\gamma$ \citep{kirkpatrick06}.  The gravity classification of the {\sc int-g} spectrum (2M~1547-24) is described in Section 4.3.  For comparison, the spectrum of a field M9.5 \citep[BRI 0021-0214;][]{cushing05} is displayed.  Young, low-gravity L0s have weaker \ion{Na}{1}, \ion{K}{1} and FeH features than normal field L0 dwarfs.}
\end{figure}

\begin{figure}
\vskip 0.5in
\centerline{\includegraphics[width=16cm]{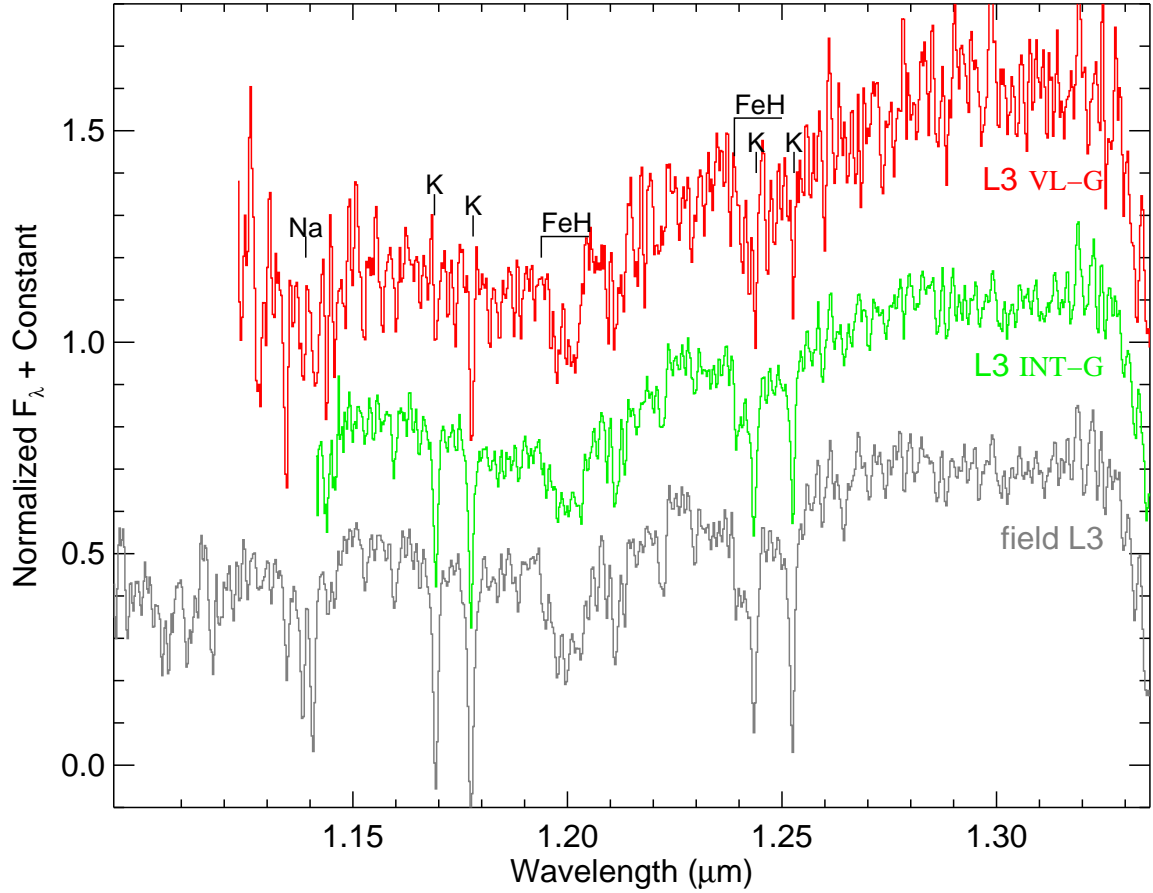}}
\vskip 2ex
\caption[Moderate-resolution spectra of L3 objects]
{\label{l3_sxd} Moderate-resolution spectra comparing the gravity (age) sensitive features in the $J$-band spectra of objects classified as L3 in the near-IR.  The L3 {\sc vl-g} spectrum (red) is G~196-3B, a companion to the 20--300~Myr old M3 star, G~196-3A.   The {\sc int-g} spectrum (green; 2M~1726+15) has an optical gravity classification of $\beta$.  For comparison, the spectrum of a field L3 \citep[2M~1506+13;][]{cushing05} is displayed.  Young, low-gravity L3s have weaker \ion{K}{1} and FeH features than normal field L3 dwarfs.}
\end{figure}

There are several hallmarks of youth (low gravity) in the 0.9--2.5~$\mu$m spectra of late-M and L dwarfs.  Gravity-sensitive features in the near-IR spectra of late-M dwarfs were originally identified by comparison of dwarf and giant spectra \citep[e.g.][]{kleinmann86,joyce98,meyer98}.  Not surprisingly, many of these same spectral features are seen in the spectra of young, late-M and early-L type brown dwarfs in star-forming regions \citep[e.g.][]{lucas01,gorlova03, mcgovern04, allers07,lodieu08}.
In the current work, we seek to identify and quantify gravity-sensitive features over a broad range in spectral type (M5--L7). Figures~\ref{m8_prz}--\ref{l3_prz} compare low-resolution spectra of objects having field gravity, intermediate gravity, and very low gravity as determined from optical spectroscopy (see Table~\ref{tbl:spt} for references).  At lower gravity, the photosphere lies at lower pressure, which has an effect on several near-IR spectroscopic features.  The FeH bands (0.99, 1.20 and 1.55~$\mu$m), \ion{Na}{1} lines (1.14 and 2.21~$\mu$m), and \ion{K}{1} (1.17 and 1.25~$\mu$m) lines are weaker in young, low-gravity objects than in older field dwarfs.   The VO band (1.06~$\mu$m) is stronger in the spectra of young objects than in older field objects.  The continuum shape of the $H$-band spectra of young objects has a distinctive ``triangular'' shape, whereas the older field object tends to display more of a ``plateau.''   The $K$-band continua of young objects (and dusty objects) have a more positive spectral slope from 2.15 to 2.25$\mu$m than seen in normal field dwarfs of the same spectral type.  Figures~\ref{m8_sxd}--\ref{l6_sxd} compare the features in moderate-resolution $J$-band spectra for objects of various ages and gravities.  At moderate resolution, the equivalent widths (EWs) of \ion{Na}{1} and \ion{K}{1} are sensitive to gravity. In the following sections, we examine each of these features and determine the utility of these youth indicators as a function of spectral type and resolution.

\begin{figure}
\vskip 0.5in
\centerline{\includegraphics[width=12cm]{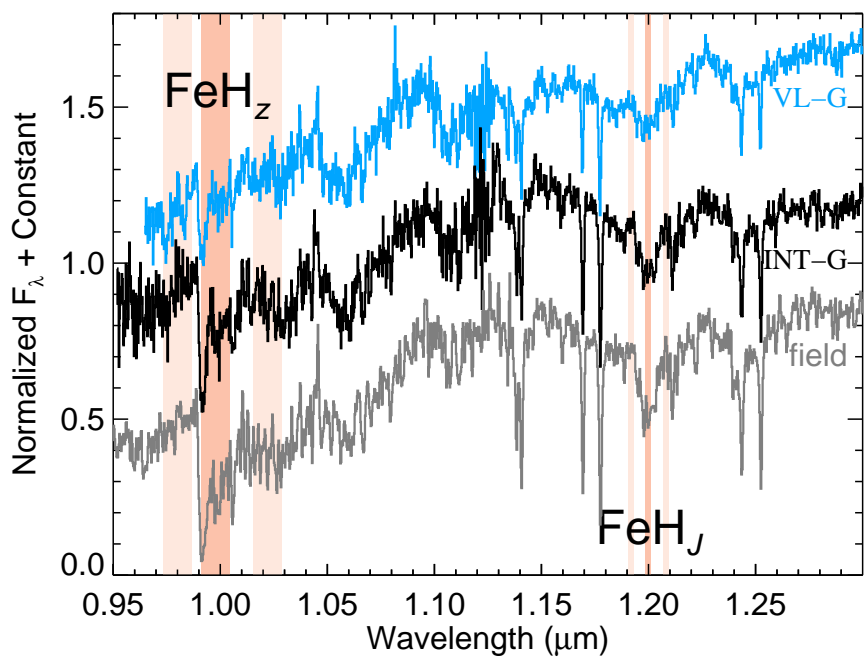}}
\vskip 2ex
\caption[Wavelength Windows for FeH Index Calculations]
{\label{fehdef} Moderate-resolution spectra showing the line (dark salmon shaded regions) and continuum (light salmon shaded regions) windows for the FeH$_z$ and FeH$_J$ indices (see Table \ref{tbl:indices} for details).  The blue spectrum is 2M~0141-46, an L0 {\sc vl-g}, which has an optical gravity classification of $\gamma$ \citep{kirkpatrick06}.  The black spectrum is the L0 {\sc int-g}, 2M~1547-24.  For comparison, the spectrum of a field M9.5 \citep[BRI 0021-0214;][]{cushing05} is displayed.}
\end{figure}

\begin{figure}
\vskip 0.5in
\centerline{\includegraphics[width=12cm]{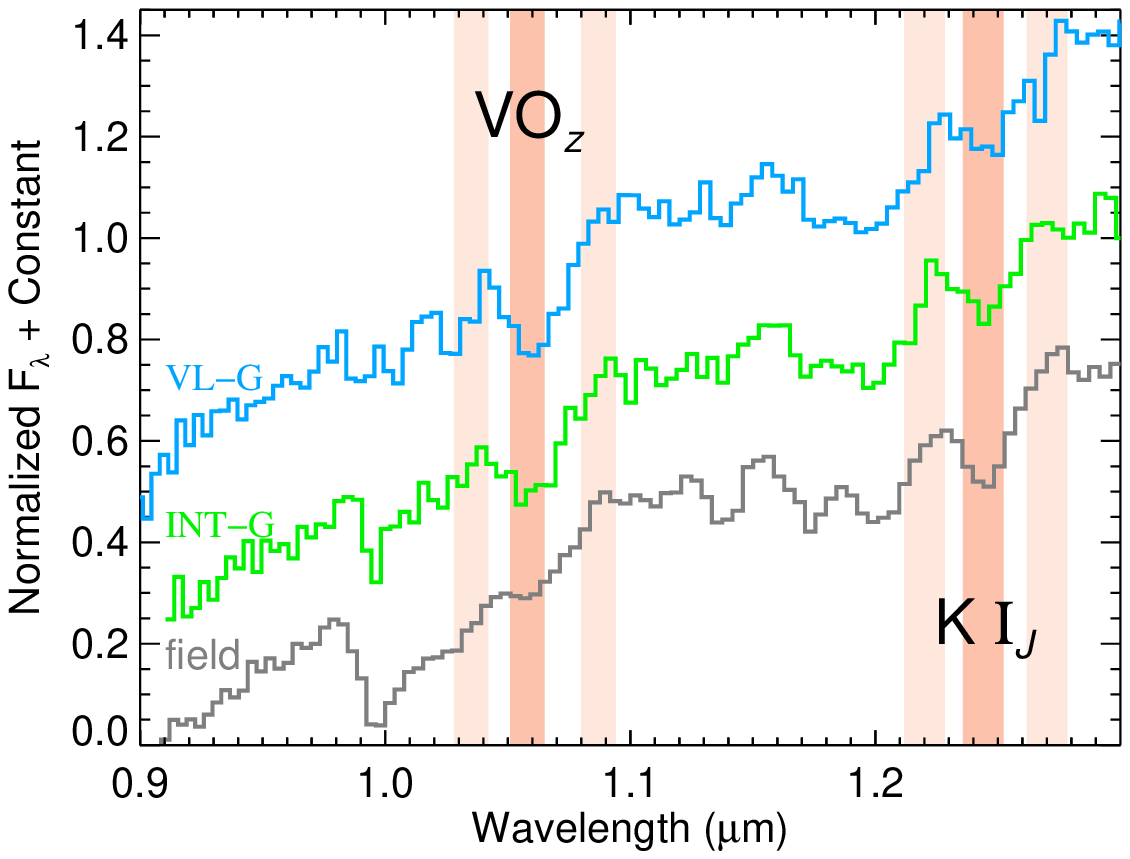}}
\vskip 2ex
\caption[Wavelength Windows for VO$_z$ and \ion{K}{1}$_J$  Index Calculations]
{\label{vo_kidef} Low-resolution spectra showing the line (dark salmon shaded regions) and continuum (light salmon shaded regions) windows for the VO$_z$ and \ion{K}{1}$_J$ indices (see Table \ref{tbl:indices} for details).  The blue spectrum is the L3 {\sc vl-g} object, 2M~2208+29, which  is classified as L3$\gamma$ in the optical.  The green spectrum is the L3 {\sc int-g} object, 2M~1726+15, which is classified as L3$\beta$ in the optical \citep{cruz09}.  The field dwarf (gray; 2M~1506+13) is an L3 standard from \citet{burgasser07c}. }
\end{figure}

\begin{figure}
\vskip 0.5in
\centerline{\includegraphics[width=12cm]{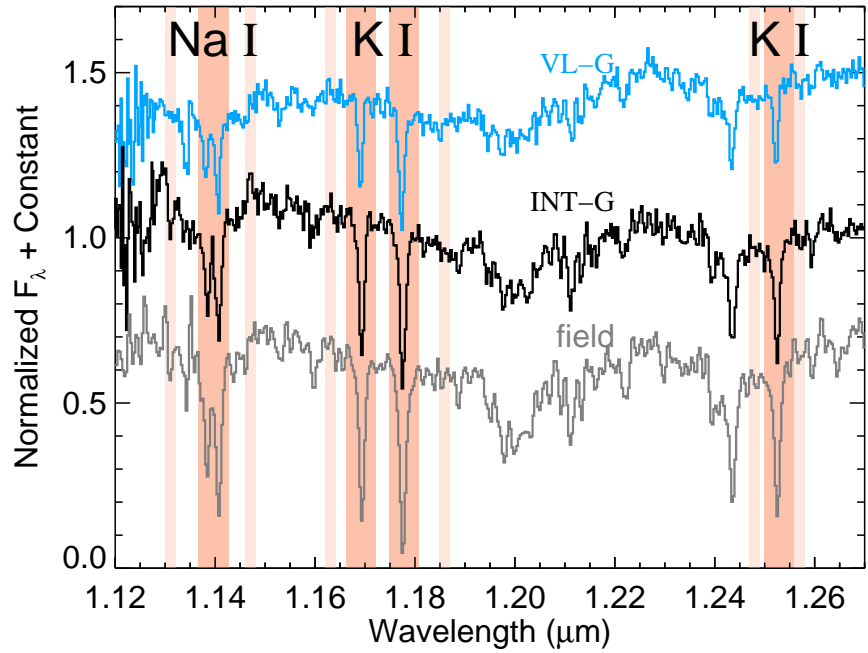}}
\vskip 2ex
\caption[Wavelength Windows for Alkali Line EW Calculations]
{\label{ewdef} Moderate-resolution spectra showing the line (dark salmon shaded regions) and continuum (light salmon shaded regions) windows for alkali line equivalent width calculation (see Table \ref{tbl:ews} for details).  The blue spectrum is 2M~0141-46, an L0 {\sc vl-g}, which has an optical gravity classification of $\gamma$ \citep{kirkpatrick06}.  The black spectrum is the L0 {\sc int-g}, 2M~1547-24.  For comparison, the spectrum of a field M9.5 \citep[BRI 0021-0214;][]{cushing05} is displayed.}
\end{figure}

\begin{figure}
\vskip 0.5in
\centerline{\includegraphics[width=12cm]{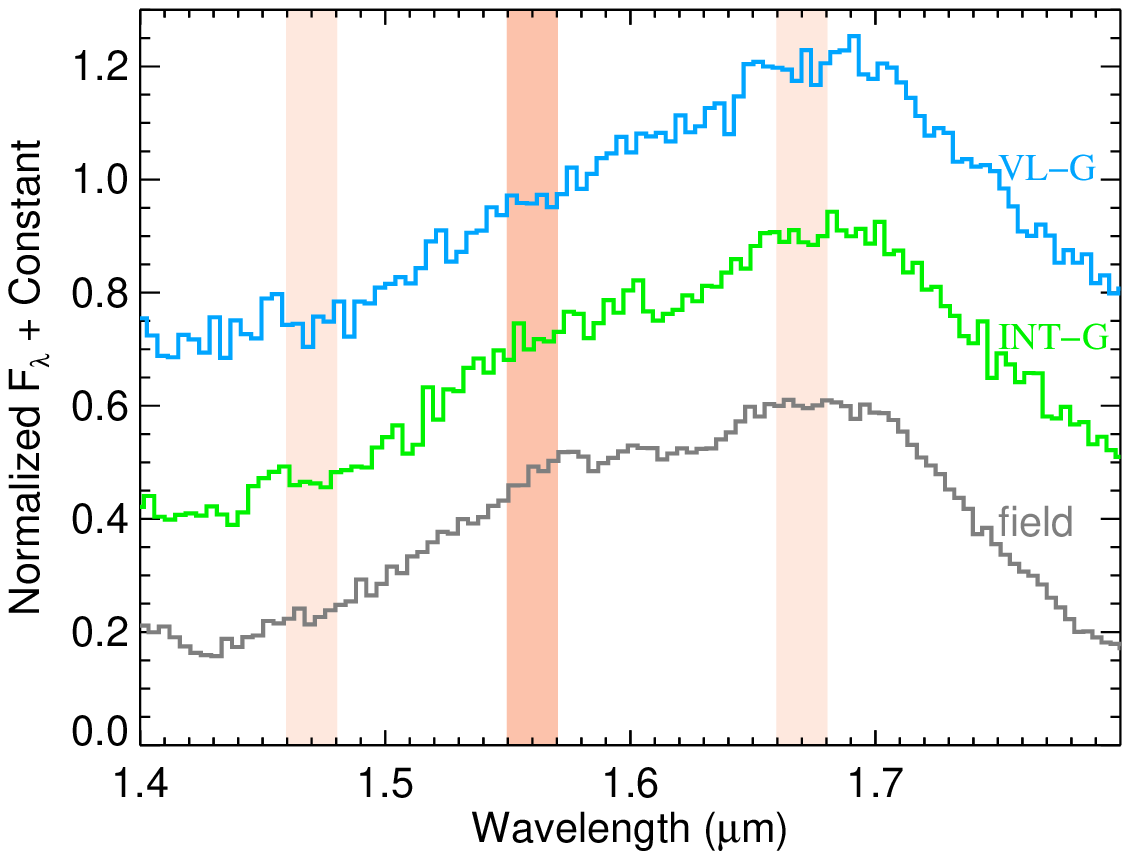}}
\vskip 2ex
\caption[Wavelength Windows for $H$-cont Index Calculations]
{\label{hcontdef} Low-resolution spectra showing the line (dark salmon shaded regions) and continuum (light salmon shaded regions) windows for the $H$-cont index (see Table \ref{tbl:indices} for details).  The blue spectrum is the L3 {\sc vl-g} object, 2M~2208+29, which  is classified as L3$\gamma$ in the optical.  The green spectrum is the L3 {\sc int-g} object, 2M~1726+15, which is classified as L3$\beta$ in the optical \citep{cruz09}.  The field dwarf (gray) is the L3 standard 2M~1506+13 \citep{burgasser07c}. }
\end{figure}

The youth of ultracool objects has typically been determined by qualitative spectroscopic comparison of the object to field dwarfs.  Here, we establish a set of spectral indices that can be used to evaluate the youth of objects in quantitative fashion.  Our basic approach to establishing these spectral indices is to center an index on a feature known to be gravity-dependent and adjust the index definition so that $\sim$10 Myr old objects in our sample (TWA members and objects with optical gravity classifications of $\gamma$) have index values that are quantitatively distinct from older field dwarfs.  Table~\ref{tbl:indices} and Figures \ref{fehdef} -- \ref{hcontdef} present our gravity-sensitive indices.

\subsubsection{FeH}

The near-IR spectra of late-M and L dwarfs contain a wealth of absorption features attributed to FeH bands \citep{mclean03, cushing05}, which also provide significant atmospheric opacity \citep{rice10a}.  The most prominent FeH features seen at low resolution are bandheads at 0.99, 1.20 and 1.55~$\mu$m.  The depth of the FeH absorption features increases steadily through the late-M to early-L spectral types, with spectral types of L2--L3 having the strongest FeH features.  For spectral types later than L3, the strengths of the FeH features decrease.  By spectral types of L7, very little FeH is discernable.  Low-gravity M and L dwarfs display much weaker FeH bands than field dwarfs of the same spectral type.

We establish the FeH$_z$ index (Table~\ref{tbl:indices} and Figure~\ref{fehdef}) which measures the depth of the 0.99~$\mu$m FeH feature.  We optimized the index to be sensitive to gravity for both moderate and low resolution spectra by making the wavelength windows for the index as wide as a single resolution element for our lowest resolution ($R \sim$ 75) spectra.  The index is calculated as follows:
\begin{equation}
{\rm index} =  \left( \frac{\lambda_{\rm{line}}-\lambda_{\rm{cont1}}}{\lambda_{\rm{cont2}}-\lambda_{\rm{cont1}}}  F_{\rm{cont2}} + \frac{\lambda_{\rm{cont2}}-\lambda_{\rm{line}}}{\lambda_{\rm{cont2}}-\lambda_{\rm{cont1}}}F_{\rm{cont1}} \right) / F_{\rm{line}}.
\end{equation}
The central wavelengths and widths of the line and continuum regions are listed in Table~\ref{tbl:indices}.  The numerator of the equation gives the expected flux at the line wavelength (based on a linear interpolation of flux in the continuum windows) if no absorption or emission were present.  $F_{\rm{cont1}}$ is the average of the spectrum (in $F_{\lambda}$ units) over a window as wide as the bandwidth listed in Table~\ref{tbl:indices}  and centered on $\lambda_{\rm{cont1}}$. $F_{\rm{cont2}}$ and $F_{\rm{line}}$ are calculated similarly.  Indices calculated using Equation~(1) will have values of one for spectra showing no FeH absorption, with higher index values indicating deeper absorption features.  Uncertainties in $F_{\lambda}$ per pixel were estimated from the rms scatter about a linear fit to wavelength versus $F_{\lambda}$ in the continuum windows.  Uncertainties in the index value were calculated assuming that the line region has the same flux uncertainty per pixel as the continuum.  We established the expected field dwarf index (black line in Figure~\ref{zindex}) and its uncertainty (gray shaded region in Figure~\ref{zindex}) from the mean and standard deviation of the index values of all field dwarfs in $\pm$ 1 subtype bins
(\eg, the field dwarf index value and uncertainty for L0 are the average and standard deviation of indices for the M9--L1 field dwarfs).  Table~\ref{tbl:prz} and Figure~\ref{zindex} present the FeH$_z$ indices calculated for our sample.  This index is sensitive to gravity for M6--L7 spectral types.  

Many of our moderate-resolution spectra do not extend to short enough wavelength to calculate the FeH$_z$ index.  
To measure the 1.20~$\mu$m FeH feature, we created the FeH$_J$ index.  At low spectral resolution, the 1.20~$\mu$m FeH feature is blended with \ion{Fe}{1}, \ion{Mg}{1} and \ion{K}{1}, thus the FeH$_J$ index is only appropriate for moderate-resolution spectra.  Table~\ref{tbl:sxd} and Figure~\ref{naew} display the FeH$_J$ indices calculated from our moderate-resolution spectra.  The 1.55~$\mu$m FeH feature contributes to the measured values of the $H$-cont index and is discussed in Section 4.2.4. 


\subsubsection{VO}

Condensation effects and higher metal hydride opacities contribute to the weaker strength of vanadium oxide (VO) bands in field dwarfs compared to low-gravity (young) ultracool dwarfs.  
We established a gravity-sensitive VO$_z$ index (Table~\ref{tbl:indices} and Figure \ref{vo_kidef}), similar to the $z$--VO index presented in \citet{cushing05}, but optimized for low-resolution spectra.  A larger value of the VO$_z$ index corresponds to deeper 1.06~$\mu$m VO absorption.  Figure~\ref{zindex} shows the index calculated for our sample, using Equation~(1).  We established the expected field dwarf index and its uncertainty in the same manner as for the FeH$_z$ index.   We find that the 1.06~$\mu$m VO feature is an excellent gravity indicator for L0--L4 dwarfs, with the index values for optically-classified L$\gamma$ dwarfs lying well above the field dwarf sequence.  Young, late-M dwarfs also show enhanced VO absorption, but the difference between young M dwarfs and field dwarfs is more subtle.  We do not use the other notable VO band (at $\sim$1.17~$\mu$m) in our analysis as this feature is blended with H$_2$O, FeH, and \ion{K}{1} features.

\subsubsection{Alkali Lines}

\ion{Na}{1} and \ion{K}{1} alkali lines are the most prominent features in the $J$-band spectra of late-M and L field dwarfs.  Pressure broadening and the condensation of opacity sources contribute to the large EWs \citep[$\sim$10~\AA;][]{cushing05} measured for these features at spectral types of L0--L6.   The \ion{K}{1} and \ion{Na}{1} lines are blended with FeH, \ion{Fe}{1}, and H$_2$O features in low resolution spectra of late-M to mid-L dwarfs, which limits the reliability of these features as age indicators at low spectral resolution.  \citet{allers07} established a gravity-sensitive index from the 1.14~$\mu$m \ion{Na}{1} line, which is appropriate for $R \gtrsim$~300 spectra.  We tested this index on the moderate-resolution spectra in our sample and found that while the index does an excellent job of distinguishing low-gravity M dwarfs from field M dwarfs, it is not as effective at determining the gravity of L dwarfs.  

We have established a \ion{K}{1}$_J$ index to measure the depth of the 1.244 and 1.253~$\mu$m \ion{K}{1} feature at low resolution.  The index is calculated using Equation~(1).  The index is contaminated by an FeH feature at 1.239~$\mu$m.  Fortunately, the trend of FeH strength with gravity is similar to the dependence of the \ion{K}{1} line depth with age.  Figure~\ref{jindex} shows the \ion{K}{1}$_J$ index, which is sensitive to gravity for spectral types of M5--L6 for low-resolution spectra.  

From moderate-resolution spectra, we can calculate the pseudo-EWs of the \ion{K}{1} and \ion{Na}{1} lines.  Table~\ref{tbl:ews} and Figure~\ref{ewdef} show the line and continuum wavelengths for our calculation of EWs.  We use line windows of 0.006~$\mu$m and continuum windows of 0.002~$\mu$m (both of which are much larger than a single resolution element for $R$=750).  We approximate the continuum in the line region from a linear fit to the flux in the continuum windows, and use the rms scatter about the fit in the continuum windows as the flux uncertainty per pixel.  We then propagate uncertainties using a Monte Carlo method to compute the uncertainty in the EW.  We calculated EWs for the \ion{Na}{1} lines centered at 1.1396~$\mu$m, and the \ion{K}{1} lines centered at 1.1692, 1.1778, 1.2437 and 1.2529~$\mu$m which are presented in Table ~\ref{tbl:sxdews} and Figures \ref{naew} and \ref{kiew}.  We do not use the 1.2437~$\mu$m \ion{K}{1} EW in our youth analysis, as it is blended with an FeH feature, which results in large uncertainties (see Figure \ref{kiew}).  
The $J$-band \ion{Na}{1} and \ion{K}{1} EWs show similar dependence on spectral type and gravity, with TWA members have substantially weaker absorption than field dwarfs.

\subsubsection{$H$-band Continuum Shape}

\begin{figure}
\vskip 0.5in
\centerline{\includegraphics[width=16cm]{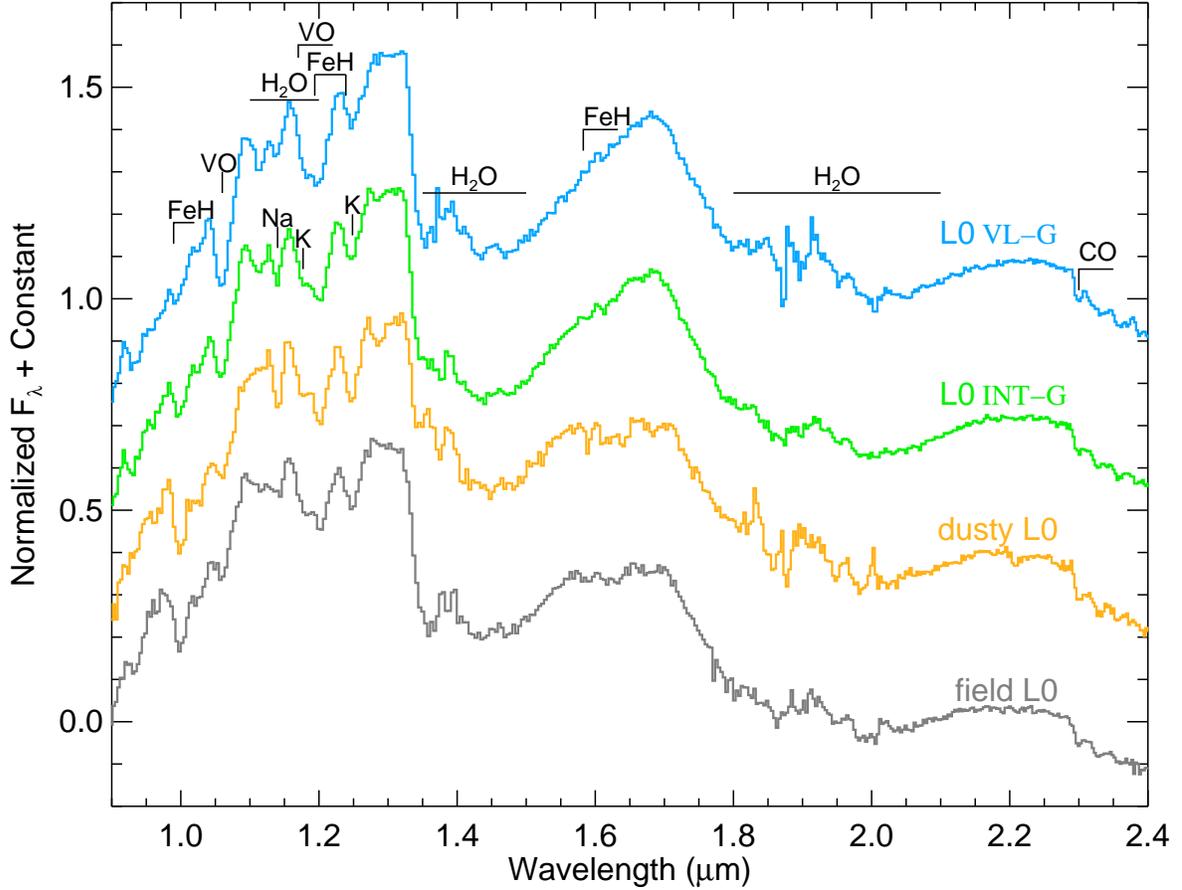}}
\vskip 2ex
\caption[Low-resolution spectra of L0 objects]
{\label{l0_prz} Low-resolution spectra comparing the gravity (age) sensitive features for objects classified as L0 in the near-IR.  From their optical spectra, the {\sc vl-g} object (blue; 2M~0141-46) is classified as L0$\gamma$ and the {\sc int-g} object (green; 2M~1552+29) is classified as L0$\beta$ \citep{cruz09}.  The spectrum of the dusty object (orange; 2M~1331+34) is from \citet{kirkpatrick10}.  The field dwarf (gray) is the L0 standard 2M~0345+25 \citep{burgasser06}.  FeH, \ion{Na}{1}, \ion{K}{1}, features are weaker at lower gravities and VO is stronger.  The $H$-band continuum shows a distinct triangular shape at low gravities.}
\end{figure}

\begin{figure}
\vskip 0.5in
\centerline{\includegraphics[width=16cm]{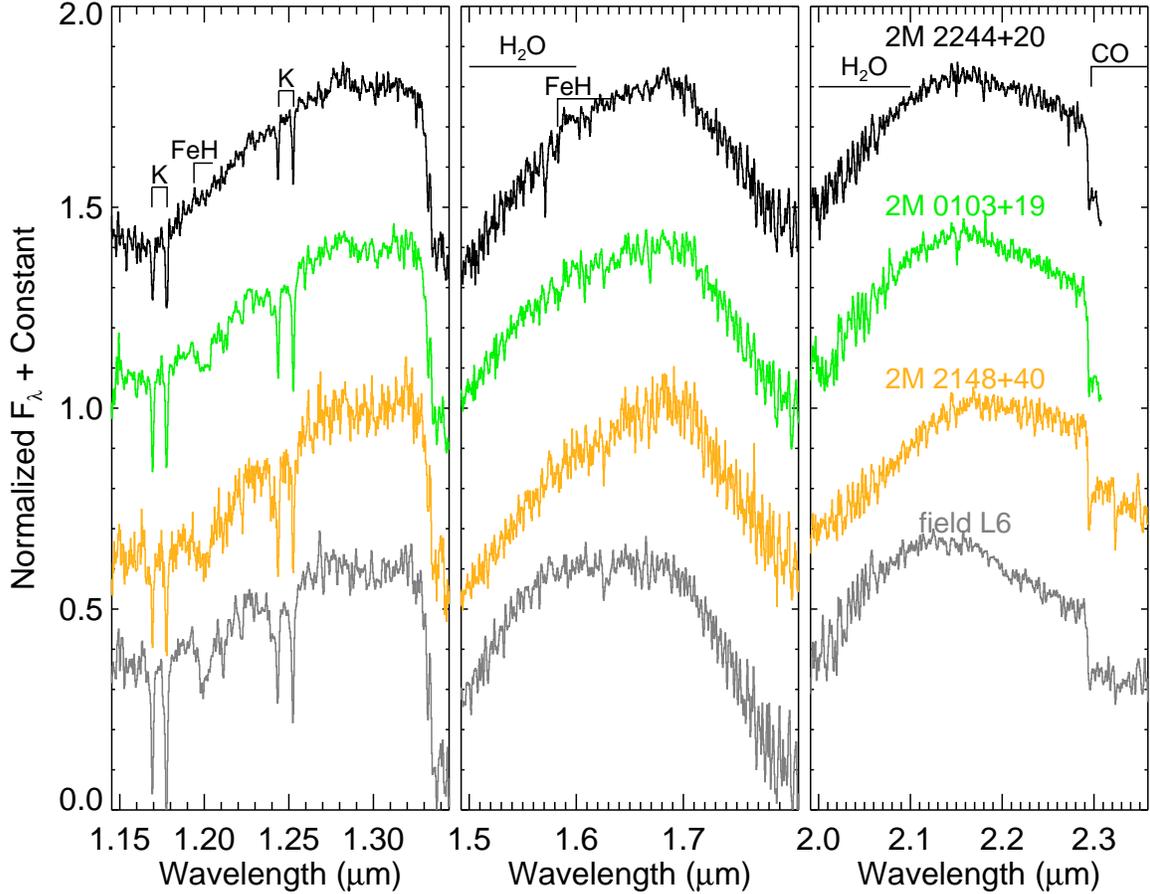}}
\vskip 2ex
\caption[Moderate-resolution spectra of L6 objects]
{\label{l6_sxd} Moderate-resolution spectra comparing low-gravity, dusty and field L6 dwarfs.  The field dwarf L6 spectrum (gray) is 2M~1515+48 from \citet{cushing05}.  The $J$, $H$, and $K$ bands are plotted separately and normalized by the mean flux from 1.27--1.32, 1.65--1.72, and 2.15--2.25~$\mu$m, respectively.  We classify 2M~2244+20 (black) as {\sc vl-g} and 2M~0103+19 (green) as {\sc int-g}.  2M~2148+40 \citep[orange;][]{looper08} is a dusty L6 that does not show signatures of youth in its kinematics or optical spectrum.  Based on its IR spectrum, we would classify 2M~2148+40 as {\sc fld-g}, as its alkali line EWs and FeH features are consistent with the L6 field dwarf.  The $H$-band continuum shape of 2M~2148+40, however, appears to have the triangular shape indicative of low-gravity.  Thus, the $H$-band continuum shape (and our $H$-cont index) alone can not reliably distinguish between low-gravity and dusty photospheres.}
\end{figure}

A hallmark of youth seen in the near-IR spectra of late-M and L dwarfs at low spectral resolution is a triangular $H$-band continuum shape (Figure~\ref{l0_prz}).  Though both very low and intermediate-gravity objects have a triangular shape compared to field dwarfs, intermediate-gravity objects appear to have a ``shoulder'' at $\sim$1.57~$\mu$m, likely due to a combination of increased FeH absorption and H$_2$ collision induced absorption \citep{borysow97}.  As noted by \citet{bowler12}, the $H$-band continuum shape for intermediate-age ($\sim$50--150~Myr) M9 dwarfs is less triangular than that of a $\sim$12~Myr old TWA M9 dwarf.  We note that the $K$-band continuum shape also appears to be sensitive to gravity (Figures \ref{l0_prz} and \ref{l3_prz}), with low-gravity objects having slightly more positive 2.15--2.25~$\mu$m spectral slopes than field dwarfs.  The differences in $K$-band spectral shape with gravity, however, are subtle, and dusty objects have $K$-band continuum shapes that are similar to young objects (Figure \ref{l6_sxd}), so we do not use $K$-band continuum shape in our analysis.

We establish the $H$-cont index to measure the triangular shape of the $H$-band spectrum (Table \ref{tbl:indices} and Figure \ref{hcontdef}).  Though calculated using Equation~(1), the $H$-cont index does not measure the depth of an absorption feature, but rather measures how much the shape of the blue end of the $H$-band deviates from a straight line.  For very low-gravity objects the blue side of the $H$-band is nearly a straight line, corresponding to an $H$-cont index value of $\sim$1.0.  Higher gravity objects have lower $H$-cont indices.    The $H$-cont index is sensitive to gravity for spectral types of M6--L7.  Spectra for objects earlier than M6 tend to have a fairly flat $H$-band continuum shape, so the index loses its efficacy for early to mid-M dwarfs.  The $H$-cont indices for our sample are displayed in Figure~\ref{hindex}.  Though objects classified as low gravity in the optical ($\gamma$ and $\beta$) appear to have low gravity using the $H$-cont index, older ``dusty'' field dwarfs also show a hint of low gravity in this index (\eg, Figure \ref{l6_sxd}).  \emph{Thus the continuum shape of the $H$-band is not the most reliable gravity indicator, particularly for intermediate gravities, and should be used in combination with other gravity-sensitive features.}



\subsection{A Quantitative Near-IR Gravity Classification Scheme}

Having created a set of gravity-sensitive indices, our goal is to utilize the indices to quantitatively classify the gravities of objects in our sample.  An optical gravity classification scheme has been developed for L dwarfs \citep{cruz09}.   Our near-IR gravity classification is designed to provide classifications that are consistent with the established optical classification system.

\begin{figure}
\vskip 0.5in
\centerline{\includegraphics[width=9cm]{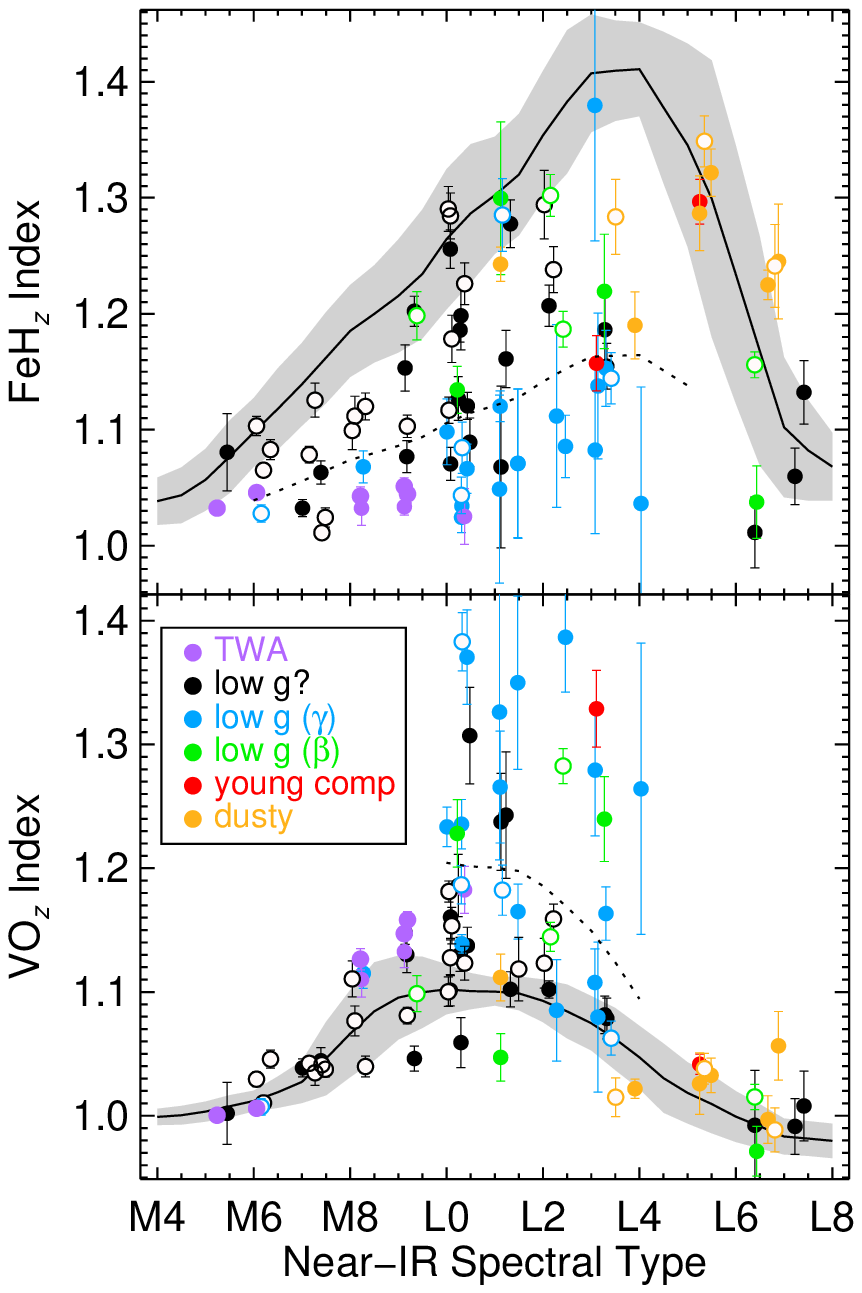}}
\vskip 2ex
\caption[$z$-band Gravity Sensitive Indices]
{\label{zindex} Gravity-sensitive indices calculated from $z$-band spectra.  The solid black line and shaded gray region show the average and standard deviation of index values as a function of spectral type for normal field dwarfs \citep{burgasser10, cushing05, geissler11, kirkpatrick10}.  Purple points represent members of the TW~Hydra moving group ($\sim$10 Myr old).  Objects in our sample with an optical gravity classification of $\beta$ are displayed as green points and those having an optical classification of $\gamma$ are displayed as blue points \citep{kirkpatrick08, cruz09, rice10, kirkpatrick10}.  Black points show objects in our sample having no optical gravity classification.  Red points represent young companions to stars.  Objects having normal gravities but thought to have unusually dusty photospheres are displayed as orange points.  Filled circles show index values calculated from low resolution ($R \approx$100) spectra, and open circles show values calculated from moderate resolution ($R \approx$750--2000) spectra.  A gravity score of 1 is assigned to objects having an index value more than 1$\sigma$ away from the field dwarf sequence.  The dotted lines show the boundary for objects to be assigned a score of 2 (rather than 1).  }
\end{figure}

\begin{figure}
\vskip 0.5in
\centerline{\includegraphics[width=9cm]{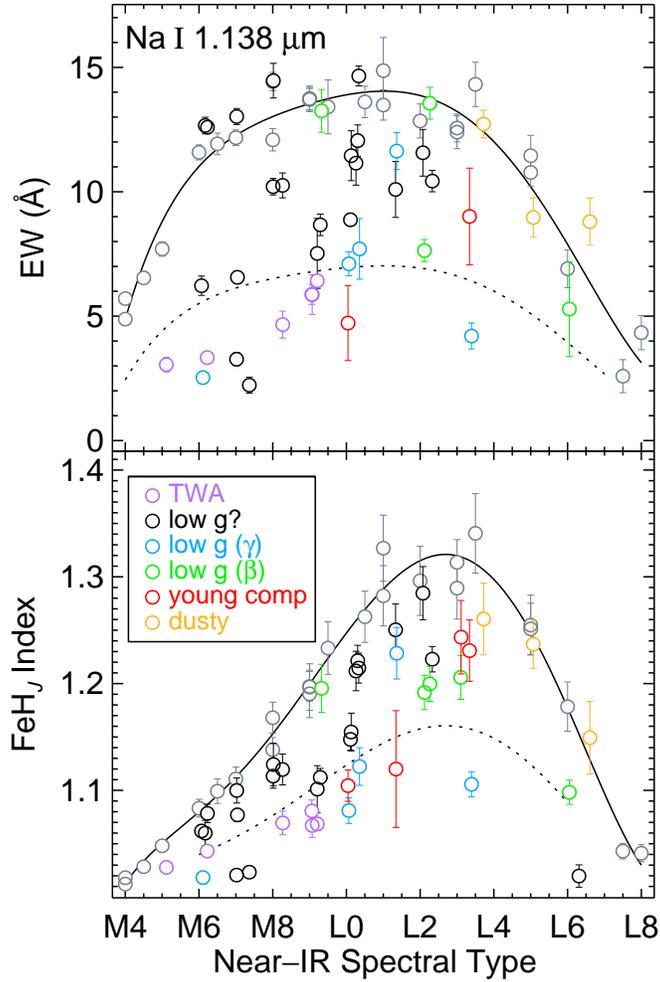}}
\vskip 2ex
\caption[Alkali Line Equivalent Widths]
{\label{naew} \ion{Na}{1} (top) and line EWs and $J$-FeH indices measured from moderate-resolution spectra.  Symbols are the same as those used in Figure~\ref{zindex}.  Gray circles show our calculated EWs for normal field dwarfs \citep{cushing05}.}
\end{figure}

\begin{figure}
\vskip 0.5in
\centerline{\includegraphics[width=9cm]{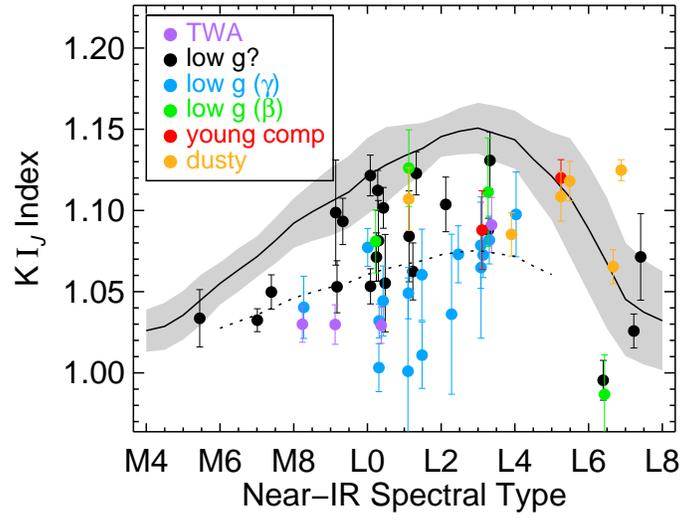}}
\vskip 2ex
\caption[$J$-band Gravity Sensitive Index]
{\label{jindex} Gravity sensitive index used to measure the depth of the 1.25~$\mu$m \ion{K}{1} doublet.  Symbols are the same as those used in Figure~\ref{zindex}.}
\end{figure}

\begin{figure}
\vskip 0.5in
\hskip -1.5in
\centerline{\includegraphics[width=3in]{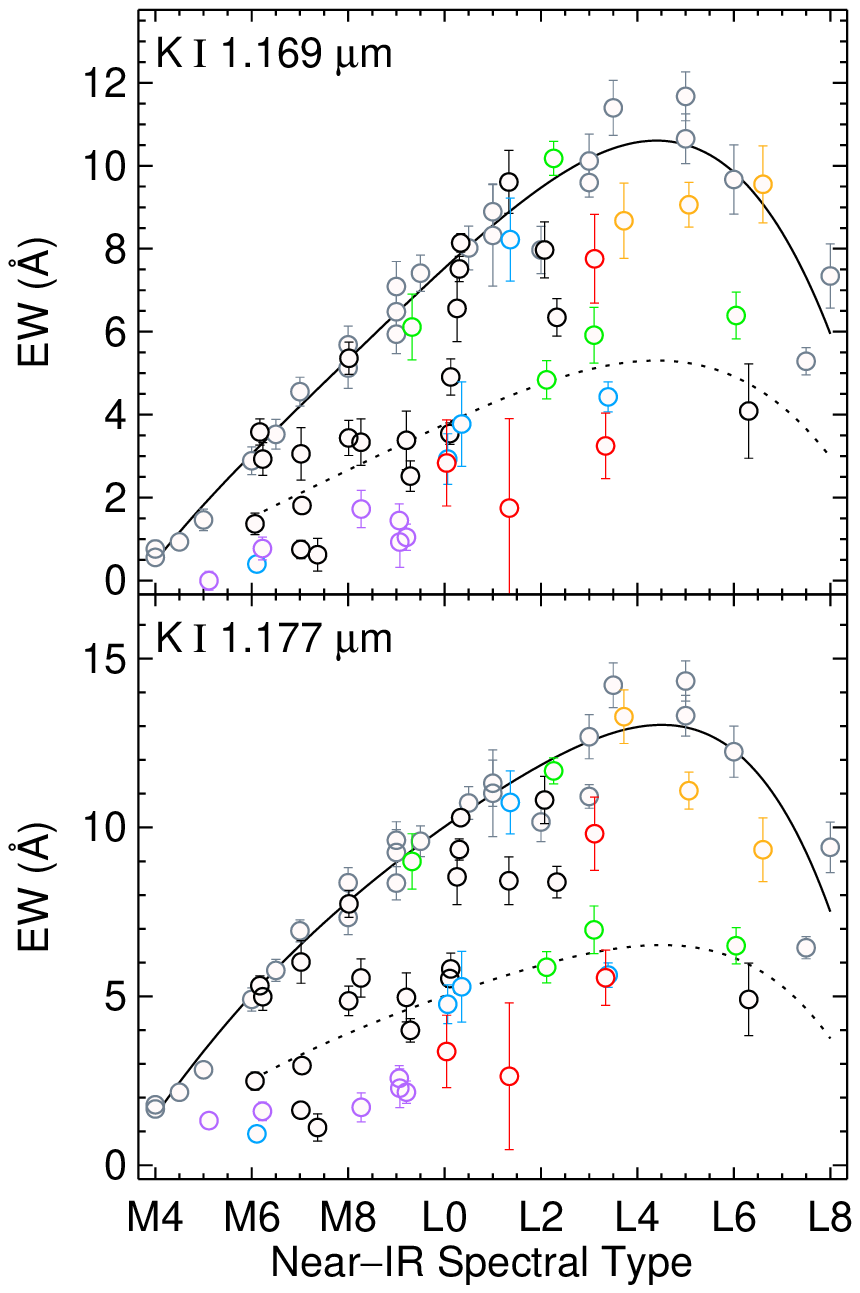}}
\vskip -4.5 in
\hskip +1.5in
\centerline{\includegraphics[width=3in]{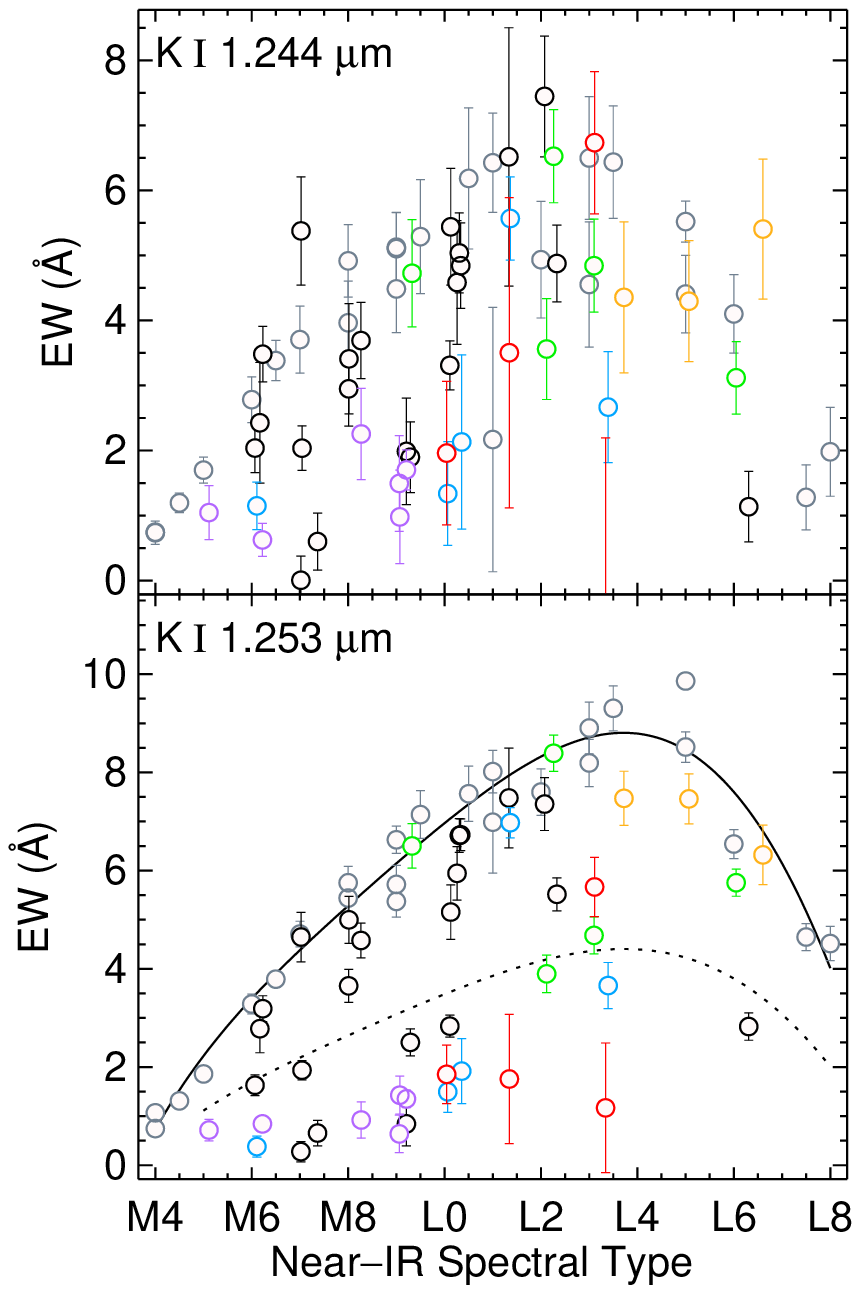}}
\vskip 2ex
\caption{\label{kiew} \ion{K}{1} equivalent widths measured from moderate-resolution spectra.  Symbols are the same as those used in Figure~\ref{zindex}.  Gray circles show the EWs calculated for normal field dwarfs from \citet{cushing05}.}
\end{figure}

\begin{figure}
\vskip 0.5in
\centerline{\includegraphics[width=9cm]{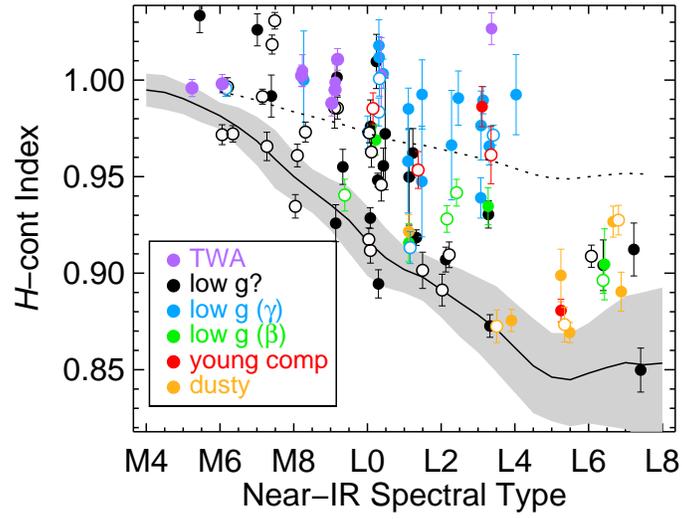}}
\vskip 2ex
\caption[$H$-band Gravity Sensitive Index]
{\label{hindex} Gravity-sensitive index measuring the continuum shape of the $H$ band. Symbols are the same as those used in Figure~\ref{zindex}.}
\end{figure}

For each index and EW we determine if the value of the index indicates low-gravity by comparison to the behavior seen in field dwarfs as function of SpT.  For the FeH$_z$, VO$_z$, \ion{K}{1}$_J$ and $H$-cont indices to indicate low gravity (a score of 1), the index value for an object must lie more than 1 $\sigma$ from the scatter in the field dwarf sequence at the object's near-IR spectral type (i.e. the object's index and error bars must lie outside of the gray-shaded regions in Figures \ref{zindex}, \ref{jindex} and \ref{hindex}).  For the FeH$_J$ index, \ion{K}{1} and \ion{Na}{1} EWs, we do not have a large enough sample of field objects observed at moderate spectral resolution to calculate the field sequence and its uncertainty as for the lower resolution indices.  Instead, we fit a fourth-order polynomial to the measured indices/EWs of the field dwarf spectra and use the scatter about the fit as the uncertainty.
The necessary index values for an object to be considered low gravity are presented in Tables \ref{tbl:indexbounds} and \ref{tbl:ewbounds}.  For each index and EW, we also create a dividing line (displayed as dashed lines in Figures \ref{zindex}--\ref{hindex})  which delineates a strong indication of low-gravity (a score of 2).  For EWs, the dividing lines were chosen to be 50\% of the field dwarf sequence.   For indices, dividing lines were set to delineate feature strengths that are a  fraction, $\alpha$, of the features in the field dwarf sequence. The dividing lines are calculated as $(({\rm field~sequence} - 1)* \alpha + 1)$, where $\alpha$ is a scale factor chosen so that the dividing line roughly separates objects having optical classifications of $\beta$ and $\gamma$.  

  We determine a gravity score for each of four indicators: FeH, VO,
  alkali lines, and $H$-band continuum shape. Following the same order,
  the scores are presented in Tables~\ref{tbl:prz} and \ref{tbl:sxd}.
  For the low-resolution spectra in our sample, we use the following
  approach to assigning scores.

\begin{itemize}

\item The gravity scores for FeH, VO, alkali line depth and $H$-band
  continuum are taken from the scores of the FeH$_z$, VO$_z$,
  \ion{K}{1}$_J$ and $H$-cont indices, respectively
  (Table~\ref{tbl:prz}). 

\item A score of 0 is given if an object's index or EW is consistent with the field dwarf sequence.

\item A score of 1 means the index indicates low gravity, with the
  index lying at least 1~$\sigma$ away from the field dwarf index and
  uncertainty. 

\item A score of 2 means the index strongly indicates low gravity and is
  farther from the field dwarf sequence than the dashed lines in Figures
  \ref{zindex}--\ref{hindex}. 

\item Scores of ``?'' are given if an index hints at low gravity, but the
  uncertainty in the calculated index is too large (i.e., the index lies
  outside of the gray shaded regions shown in Figures \ref{zindex}--\ref{hindex}, but the error bars overlap). 

\item Scores of ``n'' are assigned if either the spectrum does not fully
  cover the wavelength range of the index or the index is not
  gravity-sensitive for the object's spectral type.

\end{itemize}

For our moderate-resolution spectra, we determine the gravity scores in
a similar manner as for low-resolution spectra, but include 
  the measurements for the FeH$_J$ index and the \ion{Na}{1} and
  \ion{K}{1} EWs when determining the gravity scores. Here is the
  approach.

\begin{itemize}

\item The gravity scores for VO and $H$-band continuum are taken from
  the scores of the VO$_z$ and $H$-cont indices, respectively. 

\item The gravity score for FeH is assigned based on the FeH$_z$ and
  FeH$_J$ index scores. If either of the FeH$_z$ and FeH$_J$ index
  scores are 1, an FeH gravity score of 1 is assigned. If either index
  has a score of 2, an FeH gravity score of 2 is assigned. 

\item The gravity score for alkali lines is assigned based on the
  \ion{Na}{1} and three \ion{K}{1} line EWs.  We determine the alkali line score from the mean of the individual line EW scores rounded to the nearest integer.  

\end{itemize}

For a given object, the scores for the individual
  absorption species are usually in reasonable but not exact agreement,
  reflecting the underlying scatter in the strengths of these features
  as a function of spectral type. It is valuable to be able to describe
  the overall low-gravity nature of a source, analogous to the overall
  spectral morphology that is represented by the spectral type, even
  though there can be small spectral differences. 

\begin{figure}
\vskip 0.5in
\hskip -1.5in
\centerline{\includegraphics[width=3in]{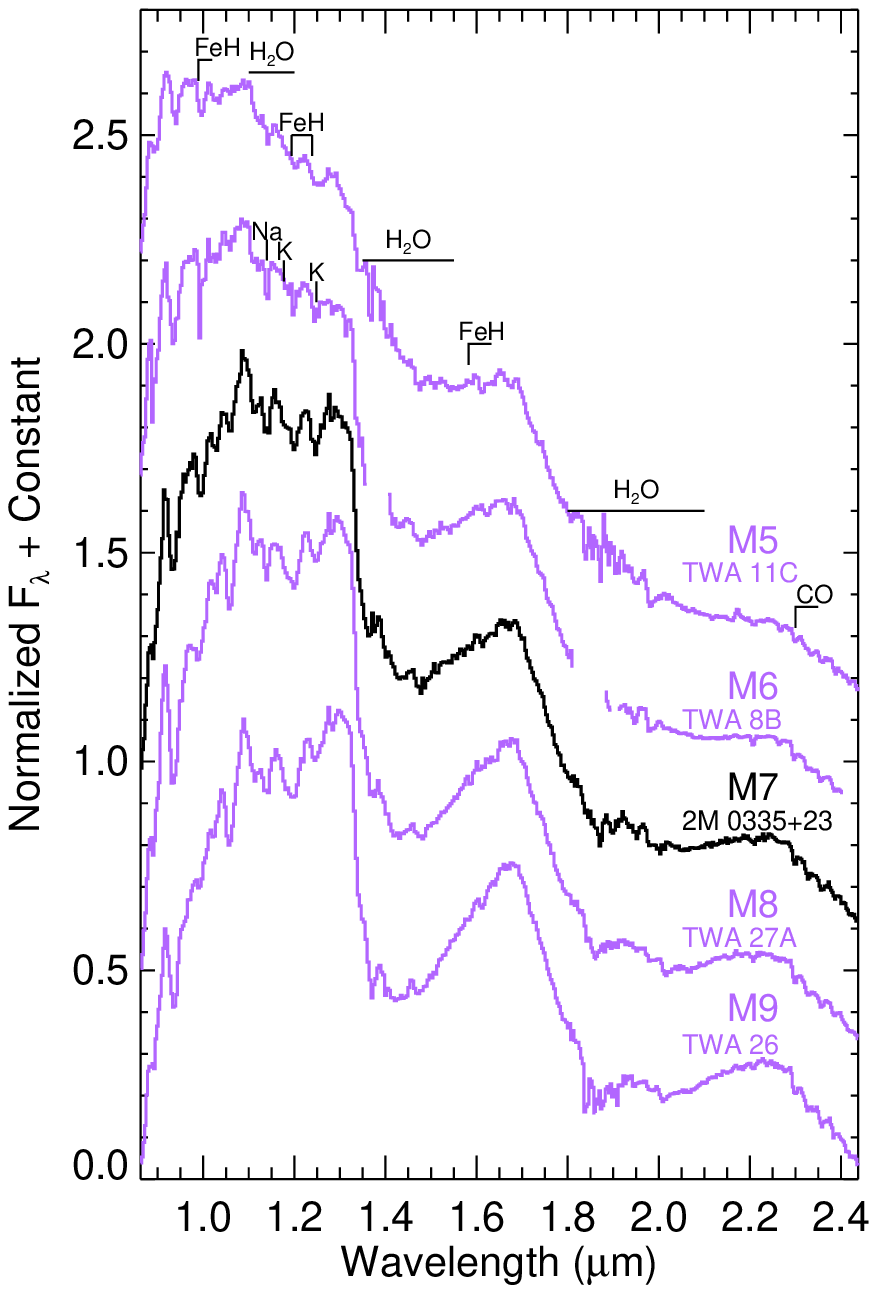}}
\vskip -4.5 in
\hskip +1.5in
\centerline{\includegraphics[width=3in]{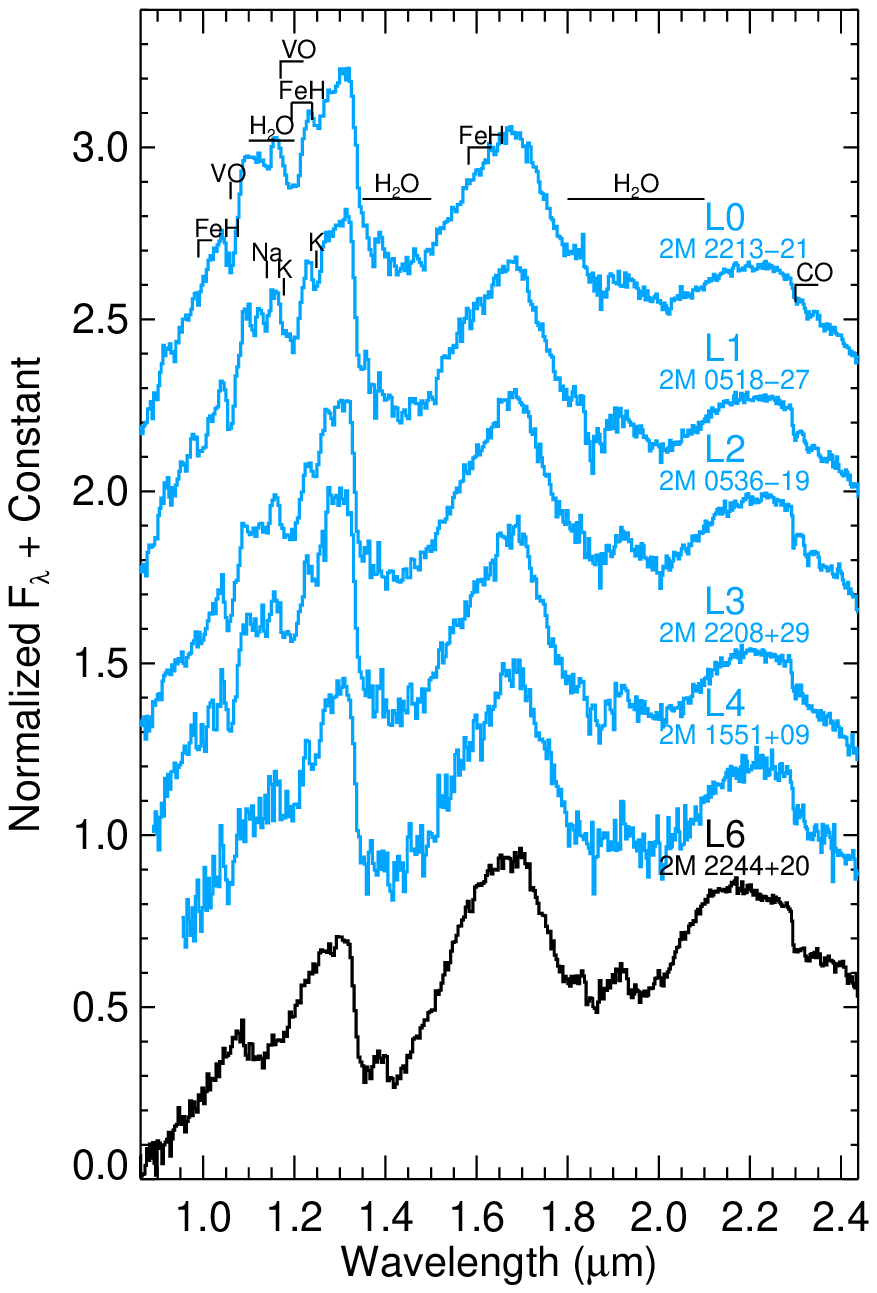}}
\vskip 2ex
\caption{\label{vlgsequence} Our sequence of field dwarfs classified as having very low gravity ({\sc vl-g}) in the near-IR.  Spectra plotted in purple are for known members of TWA.  Spectra plotted in blue are of objects having optical gravity classifications of $\gamma$.  Objects plotted in black have no available optical gravity classification.}
\end{figure}

  Desiring an objective and automatic approach, we determine an overall
  gravity classification from the median of the object's gravity scores.
  Gravity scores of n and ? are ignored for the purposes of computing
  the median. When there
  are an even number of scores, we take the average of the two values
  straddling the median, \eg, ``0101'' gives an overall score of 0.5.
  With the resulting medians, we define three near-IR gravity
  classifications.
\begin{itemize}
\item{\sc fld-g}: the object has gravity scores consistent with normal field dwarfs.  The median gravity score is $\le$0.5.
\item{\sc int-g}: the object has gravity scores consistent with intermediate gravity.  The median gravity score is 1.
\item{\sc vl-g}: the object has gravity scores consistent with very low gravity.  The median gravity score is $\ge$1.5.
\end{itemize}
The near-IR gravity classifications of our sample are presented in Table~\ref{tbl:prz}.   The Appendix gives examples of gravity classifications using our method.  Figure~\ref{vlgsequence} presents a sequence of {\sc vl-g} objects with spectral types of M5--L6.

Thirty of the near-IR spectra in our sample have gravity classifications ($\beta, \gamma$ or $\delta$) determined from optical spectra \citep{cruz09, kirkpatrick10, rice10, faherty13} where $\beta$ implies intermediate gravity and $\gamma$ and $\delta$ imply very low gravity.
Of the 21 spectra in our sample of objects having an optical classification of $\gamma$ or $\delta$, we classify 19 as {\sc vl-g}, one as {\sc int-g} (the prism spectrum of 2M~0355+11) and one (2M~1022+58) as {\sc fld-g} in the near-IR.   Thus, sources characterized as very low gravity in the optical usually have very low-gravity spectral features in near-IR as well.
Objects having optical gravity classifications of $\beta$ are generally classified as {\sc int-g} (six/nine objects). One optically classified $\beta$ object, 2M~0045+16, is classified as {\sc vl-g} in the near-IR.  Two sources, 2M~0033-15 and 2M~1022+02, having optical gravity classifications of $\beta$ show no signs of youth in any of the indicators are classified has {\sc fld-g}.  Overall, our near-IR classification system produces gravity classifications consistent with the optical system of \citet{cruz09}.

Our sample includes 16 young sources for which we have both low-resolution and moderate-resolution spectra. Our gravity classifications from both low and moderate resolution agree for all but four of these sources (LP~944-20, 2M~0355+11, 2M~2057-02 and 2M~1935-28).  For LP~944-20, both the low and moderate-resolution spectra indicate low gravity in the $H$-cont index and hint at low gravity in the VO$_z$ index (scores of ``?'').  From its moderate-resolution spectrum, the \ion{Na}{1} EW of LP~944-20 also indicates low gravity (score of 1).   The FeH$_z$ and FeH$_J$ indices as well as the \ion{K}{1} EWs of LP~944-20 are lower than the field sequence, but are not quite low enough to result in index scores of 1.   Thus, LP~944-20 is a good example of a source that is on the borderline between being classified as {\sc int-g} or {\sc fld-g}.  For 2M~2057-02, its low and moderate-resolution spectra result in classifications of {\sc int-g} and {\sc fld-g}, respectively.  The gravity scores from its moderate resolution spectrum are ??10.  Like LP~944-20, 2M~2057-02 is also a good example of a borderline source.  For 2M~1935-28, its low and moderate resolution spectra result in classifications {\sc vl-g} and {\sc int-g}, respectively.  Both the low and moderate-resolution spectra of 2M~1935-28 receive an alkali line score of 1 and an $H$-band continuum score of 2.  The difference in the gravity classifications for the two different resolutions lies in the FeH indicator.  For the low-resolution spectrum, the FeH gravity score (1) comes from the FeH$_z$ index, whereas in the moderate resolution spectrum the FeH gravity score (2) is assigned from the FeH$_J$ index.  For 2M~0355+11, both the moderate and low resolution spectra indicate low gravity, but with classifications of {\sc vl-g} and {\sc int-g}, respectively.  The low-resolution spectrum of 2M~0355+11 is fairly noisy in the $z$-band (S/N $\sim$ 10), as reflected in the high uncertainty in the FeH$_z$ index.  The \ion{K}{1}$_J$ and $H$-cont indices, as calculated from the low-resolution spectrum of 2M~0355+11, clearly indicate low gravity but do not meet the criteria for index scores of 2.  
Given the higher signal-to-noise ratio (S/N) for indices and EWs measured from moderate-resolution spectra, we adopt the classifications based on moderate-resolution spectra where available.  In general, there is good agreement between gravity determinations using low-resolution and moderate-resolution spectra which highlights the effectiveness of our indices.  

We included published spectra of dusty objects \citep{kirkpatrick10, looper08} to test if our gravity classification system correctly classifies these peculiar objects as having normal (field) gravity.  These dusty objects are plotted in Figures \ref{zindex}--\ref{hindex} as orange points.  As noted above, several of the dusty objects appear to be low gravity in the $H$-cont index.   Despite this, we classify all of the dusty objects as {\sc fld-g} when considering all of their gravity-sensitive index values.  Our index-based gravity classification can thus be used to discriminate between low-gravity and dusty objects.

\subsection{Proposed Low-gravity Spectral Standards}

The near-IR spectra of low-gravity objects can show remarkable variation in near-IR colors as well as spectral features, as shown in Figure \ref{l3_vlg}.  Our method of spectral type and gravity classification described above will determine reliable spectral types and gravities, without bias to near-IR color or preconceived ideas of gravity.  A great deal of current work relies on the use of spectral templates for determination of spectral types, particularly for optical spectroscopy \citep[e.g.][]{west11}.  There does not currently exist a set of near-IR spectral templates for low-gravity ultracool dwarfs.  Having determined the spectral types and gravities for our sample, we propose a set of  {\sc vl-g} spectral standards.  We note that these standards can be used as spectral templates, but recommend comparison to individual features, as determining best-fit templates to the entire 0.8--2.5~$\mu$m spectrum can often be biased toward fitting the closest near-IR color.  We selected possible spectral standards using the following criteria.
\begin{enumerate} 
\item Proposed low-gravity standards should have near-IR spectral types that are identical or similar to their optical spectral types.  This assures that an object classified using our proposed near-IR spectral standards will, in general, have a similar optical spectral classification.
\item Standard spectra should show a strong indication of low-gravity (a score of 2) in the majority of our gravity-sensitive indices.  Our proposed standards have firm {\sc vl-g} classifications.
\item Preference is given to objects that are known members of young kinematic groups (see Section 4.5.1) or have optical gravity classifications of $\gamma$, as this additional information helps to more fully understand their properties.
\item Preference is given to objects that have high S/N spectra, to allow for easy comparison to future work.
\end{enumerate}
Figure~\ref{vlgsequence} presents a proposed sequence of {\sc vl-g} spectral standards. Our proposed spectral standards show a smooth progression of spectral features in the $J,H,$ and $K$ windows, as well as an overall trend toward redder $J-K$ colors at later spectral types. We note that no L5 type object met the criteria listed above, so this subtype is lacking in our sequence.  

\begin{figure}
\centerline{\includegraphics[width=3in]{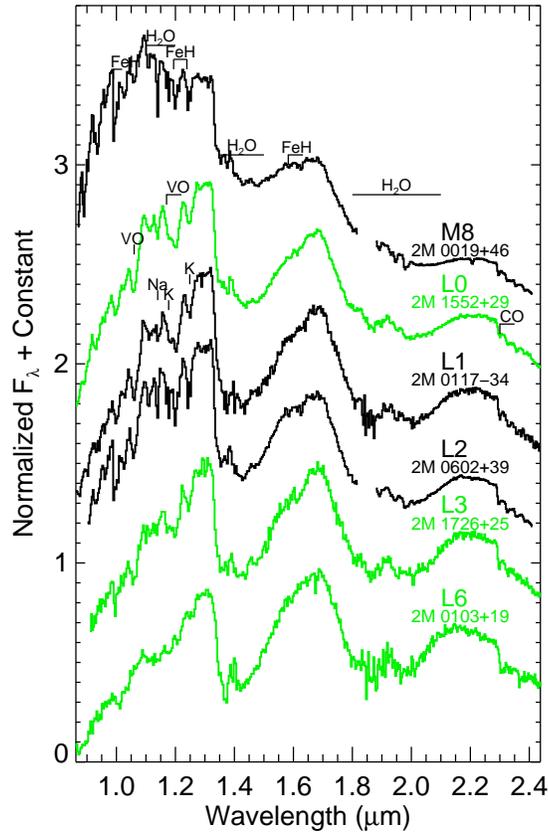}}
\vskip 2ex
\caption{\label{intsequence} A sequence of field dwarfs classified as having intermediate gravity ({\sc int-g}) in the near-IR.  Spectra plotted in green are for objects with optical gravity classifications of $\beta$.  Objects plotted in black have no available optical gravity classification.}
\end{figure}

In general, we find that objects classified as {\sc int-g} show more variation in their gravity-sensitive features at a given spectral type than objects classified as {\sc vl-g}.  Our sample also consists of fewer {\sc int-g} objects than {\sc vl-g} objects.  Thus, determining a full sequence of {\sc int-g} standards would be premature at this time.  Figure~\ref{intsequence} shows spectra for objects that meet the criteria to be {\sc int-g} standards.  Spectral types of M5--M7, M9, and L4--L5 are not represented, as suitable standards are not available in our sample.

\subsection{The Ages of Low-gravity Ultracool Dwarfs}

Determining how our gravity classifications correspond to a specific age is difficult.  In addition, the age sensitivity of our indices could be dependent on spectral type.  As shown in the previous section, our near-IR gravity classifications of {\sc vl-g} and {\sc int-g} are consistent with optical classifications of $\gamma$ and $\beta$, respectively.  Based on analysis of gravity sensitive features in optical spectra, \citet{cruz09} estimate the ages for $\gamma$ and $\beta$ classifications to be log(age (yr))~$\approx$~7 and log(age (yr))~$\approx$~8.  We can examine age estimates for a number of sources in our sample and determine the rough ages that correspond to our gravity classifications.  Table~\ref{tbl:ages} summarizes the age estimates for our sample.

\subsubsection{Young Kinematic Group Members}
\begin{figure}
\vskip 0.5in
\centerline{\includegraphics[width=16cm]{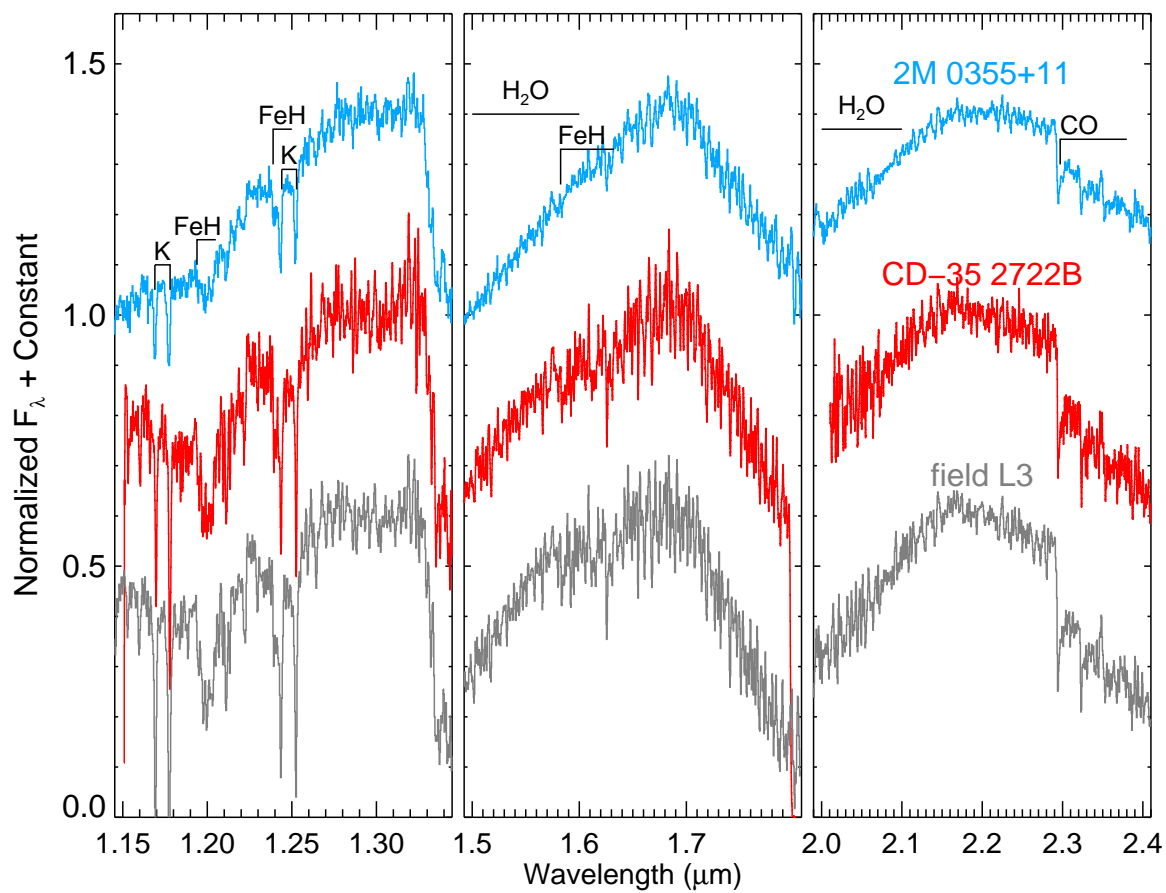}}
\vskip 2ex
\caption[Spectra of L3 AB Dor members]
{\label{l3_abdor} Moderate-resolution spectra comparing 2M~0355+11 (blue) and CD-35~2722B (red), both of which are likely members of the AB Doradus moving group \citep{liu13, wahhaj11}.  A field L3 dwarf \citep[gray;][]{cushing05} is shown for comparison.  The $J$, $H$, and $K$-bands are plotted separately and normalized by the mean flux at 1.27--1.32, 1.65--1.72, and 2.15--2.25~$\mu$m, respectively.  Membership in AB~Dor implies that 2M~0355+11 and CD-35~2722B are the same age.  Yet, based on their near-IR spectra, they are classified as having different gravities, {\sc vl-g} and {\sc int-g}, respectively.}
\end{figure}

A number of objects in our sample have been tied to young kinematic groups, which allows us examine our gravity classifications for objects of known age.  Based on their calculated index values and gravity scores, the TWA objects in our sample are all classified as {\sc vl-g}, including the L3 planetary-mass companion TWA~27B (a.k.a. 2M1207b).  This is in good agreement with the young age of TWA \citep[$\sim$8--12 Myr; ][]{torres08,mentuch08}.  One object in our sample, 2M~0608-27 \citep{rice10}, has been recently linked to the $\sim$10~Myr old $\beta$~Pictoris \citep{torres08} moving group.   2M~0608-27 is characterized by our indices as L0 {\sc vl-g}, which agrees with its possible membership in $\beta$~Pictoris.
Two objects in our sample, GSC 08047B \citep{chauvin05a} and AB PicB \citep{chauvin05b}, are companions to members of the $\sim$30~Myr old Tuc--Hor Association.   Both  GSC 08047B (L1) and the AB PicB (L0) are classified as {\sc vl-g}, as is 2M~0141-46 (L0 {\sc vl-g}), a candidate Tuc--Hor member \citep{kirkpatrick10}.  Complicating the possible age for our {\sc vl-g} classifications, 2M~0355+11 (L3 {\sc vl-g}) has a measured space motion that is consistent with membership in the $\sim$100 Myr old AB~Doradus moving group \citep{liu13}.   Interestingly, CD-35 2722B, also an AB~Dor member \citep{wahhaj11}, has only very subtle signatures of low gravity and is classified as an L3 {\sc int-g}.    Figure~\ref{l3_abdor} compares the spectra for these two objects.  Overall, it appears that an infrared classification of {\sc vl-g} corresponds to an age of $\sim$10--30 Myr, in agreement with the age estimate for optically classified $\gamma$ objects \citep[log(age(yr))~$\approx$~7;][]{kirkpatrick10}.  It is important to note, however, that older objects (e.g., 2M~0355+11) can also display very low-gravity spectral signatures.

2M~0314+16 and 2M~1705-05 are candidate members of the $\sim$500~Myr old \citep{king03} Ursa Majoris moving group \citep{seifahrt10} and neither show signatures of low gravity.  As mentioned in Section 4.3, LP 944-20 is classified as L0 {\sc fld-g} but is on the borderline for {\sc int-g} classification.  LP 944-20 is thought to be a member of the $\sim$200~Myr old Castor moving group \citep{ribas03}.  Thus, though an absolute upper limit to the ages of our {\sc int-g} sources is difficult to determine from young moving group members, we appear to be sensitive to ages $\lesssim$200 Myr.

\subsubsection{Objects with Lithium}
A number of our sources have published lithium (\ion{Li}{1}) detections from optical spectroscopy.  The detection of lithium in the spectrum of a low-mass star or brown dwarf can imply two things:  (1) the object has too little mass ($\leq 65~M_{\rm{Jup}}$) to ever have burned lithium or (2) the object is young and has not yet depleted its lithium.  For M dwarfs, a \ion{Li}{1} detection provides a useful upper limit on the ages, with detections at later spectral types corresponding to an older upper limit on the age \citep{chabrier96}.  For L dwarfs however, young objects can have weaker \ion{Li}{1} than field dwarfs \citep{kirkpatrick08} due to lower surface gravity (i.e., a similar effect as seen for \ion{Na}{1} and \ion{K}{1}).  For late L dwarfs, lithium is expected to form molecular species and \ion{Li}{1} should not be seen in their spectra.
\citet{shkolnik09} detected strong \ion{Li}{1} absorption in the optical spectra of 2M~0557-13 (M7 {\sc vl-g}) and 2M~0335+23 (M7 {\sc vl-g}) and determined upper limits to their ages of 10~Myr, consistent with our near-IR classification.  The detection of \ion{Li}{1} in the optical spectrum of SERC~296A (M6 {\sc vl-g}) has a limit on its age of $\lesssim$200~Myr \citep{thackrah97}, which is consistent with its classification.  Both 2M~0019+46 and 2M~1411-21 are 
M8 {\sc int-g} objects with \ion{Li}{1} detections \citep{reiners09}, which sets an age limit for these sources of $\lesssim$300~Myr.

\subsubsection{Companions to Young Stars}
Six of our sources are companions to stars.  Four of these objects are discussed in Section 4.5.1, as their stellar companions are known members of young kinematic groups.  Two of our sources are companions to stars for which ages can be approximated.  G~196-3B is classified as L3 {\sc vl-g} based on its near-IR indices, and has an estimated age of 20--300~Myr \citep{rebolo98, kirkpatrick01}.  The {\sc vl-g} classification of G~196--3B and spectral similarity to other {\sc vl-g} L3 objects (Figure~\ref{l3_vlg}) argues that the G~196-3 system likely falls at the low end of its estimated age range.
Gl~417B is an L4.5 companion to a G0 star.  We classify Gl~417B as L5 {\sc fld-g} but note that its spectrum shows very subtle signatures of low-gravity.  Based on a variety of indicators, \citet{kirkpatrick01} estimate an age of 80--300 Myr for Gl~417B.  Gyrochronology of Gl~417A, however, gives an age estimate of 750 Myr \citep{allers10}.   

\subsection{Notes on Selected Objects}

\subsubsection{2MASS J22443167+2043433}
With a $J-K$ color of 2.45 mag, 2M~2244+20 is one of the reddest known L dwarfs.  Its optical spectrum does not show obvious signs of peculiarity \citep{kirkpatrick08}, but its very peculiar IR spectrum has been attributed to low gravity and/or low metallicity \citep{mclean03}.  Low metallicity seems an unlikely explanation for 2M~2244+20's peculiar spectrum, as low metallicity objects tend to have much bluer $J-K$ colors \citep[e.g. the sdL7 2MASS J05325346+8246465 with $J-K$=0.26 mag;][]{burgasser07}.  Figure~\ref{l6_sxd} shows the spectrum of 2M~2244+20 (L6 {\sc vl-g}) compared to 2M~0103+19, an optical L6$\beta$ we classify in the infrared as an L6 {\sc int-g}, as well as the dusty L6.5 dwarf 2M~2148+40 \citep{looper08}.  Despite its $J-K$ color being similar to 2M~2148+40 ($J-K$=2.38 mag), the spectrum of 2M~2244+20 more closely resembles 2M~0103+29 ($J-K$=2.14 mag).   The continuum shape of 2M~2244+20's spectrum is remarkably similar to 2M~0103+29 (despite the difference in $J-K$ color for the two objects), but the \ion{K}{1} and FeH features in 2M~2244+20 are weaker implying that 2M~2244+20 has a lower gravity.  

\subsubsection{Reddened Objects}

As seen in Figure~\ref{mallspec}, 3 of the M-dwarfs in our sample (2M~0422+15, 2M~0435-14, and 2M~0619-29) have particularly red near-IR spectra for their spectral types.  
We have estimated the reddening to each of these sources by comparing their 2MASS $J-K$ colors to the photospheric colors for young objects of their spectral types \citep{luhman10}.  We find reddenings of $A_v$ = 4.6, 7.4, and 6.5 mag for 2M~0422+15, 2M~0435-14, and 2M~0619-29, respectively.  We dereddened their near-IR spectra and found that these levels of reddening do not change the spectral type or gravity determinations for these objects.

2M~0422+15 lies $\sim$10\arcdeg\ south of the Taurus star-forming region in an area that has diffuse extinction of $A_v\simeq$ 0.8--1.4~mag \citep{lombardi10} and CO emission \citep{dame01}, thus it seems plausible that it is either behind or embedded in interstellar material associated with the Taurus-Auriga region.  2M~0422+15 displays excess emission in the mid-IR (J.~Lyons et al. 2013, in preparation) indicative of a circumstellar disk, which is not particularly surprising given our classification of {\sc vl-g}. 

The diffuse extinction measured toward 2M~0435-14 is $A_v \sim$1.5 mag \citep{schlegel98} and no CO emission is detected in the immediate vicinity.  As noted by \citet{cruz03}, 2M~0435-14 lies in the direction of MBM20, a  molecular cloud 112--161~pc  away \citep{hearty00}.  \citet{cruz03} calculated a spectrophotometric distance for 2M~0435-14 of 30 pc and concluded that it could not be a member of MBM20.  2M~0435-14 does not display excess emission from a disk (J.~Lyons et al. 2013, in preparation) and is in front of the MBM20 cloud.  Thus, the source of its reddening remains unknown.

The dust map toward 2M~0619-29 indicates very low extinction ($A_V \lesssim$ 0.2) in the region \citep{schlegel98}.  2M~0619-29 has mid-IR excess emission indicative of a circumstellar disk (J.~Lyons et al. 2013, in preparation).  With a spectral type of M5, we could not assign a gravity classification to 2M~0619-29.  Qualitatively, 2M~0619-29 has low-gravity features compared to a field M5 dwarf spectrum.  This, combined with the detection of a circumstellar disk for this source make it likely to be $\lesssim$10~Myr old, and possibly reddened by its disk.

\subsubsection{2MASS J03552337+1133437}
2M~0355+11 is one of the more interesting objects in our sample.  It has one of the reddest $J-K$ colors (2.52 mag) of any known L dwarf.  It was classified as an L5$\gamma$ by \citet{cruz09}, and it is a mere 9.1$\pm$0.1~pc away \citep{liu13}.  Based on its kinematics and sky position, \citet{liu13} link 2M~0355+11 to the $\sim$100 Myr old AB~Doradus moving group.  In contrast, \citet{faherty13} determine that this object is unlikely to be an AB~Dor member, but based on a lower precision parallax measurement.  Figure~\ref{l3_abdor} shows the spectrum of 2M~0355+11 compared to CD-35~2722B, a young companion to an AB~Dor member \citep{wahhaj11}.  Despite having the same infrared spectral type (L3) and nominally the same ($\sim$100~Myr) age, the spectra of 2M~0355+11 and CD-35~2722B are quite different.  The spectrum of CD-35~2722 has very subtle hints of youth and we classify this object as {\sc int-g}, consistent with the $\sim$100~Myr age of AB~Dor.  2M~0355+11, on the other hand, has very distinct low-gravity features, and is classified as {\sc vl-g}.  We note that even if 2M~0355+11 were assigned an IR spectral type of L5 (its optical spectral type), we would still classify this object as {\sc vl-g}.  The discrepancies between the spectra of 2M~0355+11 and CD-35~2722B are puzzling.  Perhaps 2M~0355+11 and CD-35~2722 are not the same age (i.e., one of them is not a member of AB~Dor)?  If 2M0355+11 and CD-35~2722 are indeed coeval, their spectra indicate that \emph{objects of the same age may have very different spectral signatures of youth.}  

\section{Conclusions}

  We have analyzed the largest sample to date of near-IR spectra of
  young ultracool dwarfs. By comparing known young objects in our sample
  to field dwarfs, we have found that both visual and index-based
  classification works well in the near-IR, producing types that are
  well-correlated with optical spectral types. As a result, we have
  developed a method for determining near-IR spectral types that is
  gravity-insensitive.

  We also have examined our spectra for gravity (age) sensitive features
  and have constructed a set of near-IR spectral indices that measure
  the depths of VO, FeH and alkali line absorption as well as the
  $H$-band continuum shape. By comparing index measurements for young
  and old (field) ultracool dwarfs, we have created a scoring system and
  established two gravity classifications, {\sc vl-g} and {\sc int-g},
  for use with the near-IR spectra of M5--L6 objects.
  Our approach provides consistent results between optical and near-IR
  gravity classifications, with our {\sc vl-g} and {\sc int-g}
  classifications corresponding to the \citet{cruz09} optical
  gravity classifications of $\gamma$ and $\beta$, respectively.

  A subset of our sample have ages determined by kinematically linking
  them to nearby young moving groups or have limits placed on their ages
  by the detection of \ion{Li}{1} in their optical spectra. We estimate
  that objects with near-IR gravity classifications of {\sc vl-g} are
  $\sim$10--30~Myr old and those with gravity classifications of {\sc
    int-g} are $\sim$50--200~Myr old, though there are exceptions to
  these age limits (e.g., the $\approx$100~Myr AB~Dor member 2M~0355+11
  appears to be unusually low gravity). As additional kinematic
  information becomes available for young field ultracool dwarfs, more can
  be linked to young moving groups, allowing a more detailed study of
  the age-dependence of our classification system. 

\acknowledgments
We are grateful to Kimberly Aller, William Best, Brendan Bowler, Michael Cushing, Niall Deacon, Casey Deen and Geoff Mathews for obtaining some of the IRTF/SpeX observations presented here.  We thank Jackie Faherty, John Gizis, Davy Kirkpatrick, Dagny Looper, Kevin Luhman, Stanimir Metchev, Jenny Patience, Emily Rice, and Zahed Wahhaj for making their published spectra available.  
We also thank Kelle Cruz for useful discussions about the classification of low-gravity ultracool dwarfs.  We are especially grateful to Brendan Bowler and our anonymous referee for providing comments used to improve this manuscript.  This research has benefited from the M, L, and T dwarf compendium housed at DwarfArchives.org and maintained by Chris Gelino, Davy Kirkpatrick, and Adam Burgasser as well as from the SpeX Prism Spectral Libraries, maintained by Adam Burgasser at http://www.browndwarfs.org/spexprism.  This research was supported by NSF grants AST-0407441 and AST-0507833 as well as NASA Grant NNX07AI83G.

\appendix
\section{Examples of Gravity Classification}

\subsection{A Low-resolution Spectrum}

As an example of a gravity classification for a low-resolution spectrum, we will examine the spectrum of 2MASS~J17260007+1538190 (L3).  Table \ref{tbl:prz} lists the calculated indices for this spectrum.  We can determine the gravity scores for each of the calculated indices using the criteria listed in Table \ref{tbl:indexbounds}.

\begin{itemize}
\item{FeH: the FeH gravity score for low-resolution spectra is determined from the FeH$_z$ index.  The value of the FeH$_z$ index for 2M~1726+15 is $1.220 \pm 0.050$.  The index value is less than 1.163 (the requirement for an L3 to receive a score of 1 in this index) by more than one sigma, but does not meet the requirement to receive a score of 2 (FeH$_z \le 1.357$). Thus the FeH indicator is assigned a gravity score of 1.}
\item{VO: the VO gravity score is determined from the VO$_z$ index.  The VO$_z$ index of 2M~1726+15 ($1.239\pm0.035$) meets the criteria to be assigned a score of 2 (VO$_z \ge 1.097$).  Thus, the VO gravity score is 2.}
\item{Alkali Lines: the alkali score for low-resolution spectra is determined from the \ion{K}{1}$_J$ index.  The \ion{K}{1}$_J$ index value of 2M~1726+15 ($1.111 \pm 0.034$) is less than 1.135 but not by more than 1$\sigma$.  Thus, the alkali line gravity score for 2M~1726+15 is ``?''.  }
\item{$H$-band continuum shape:  the gravity score for $H$-band continuum shape is determined from the $H$-cont index.  The $H$-cont index value of 2M~1726+15 ($0.935\pm0.010$) meets the criteria for an L3 type object receive a score of 1 ($H$-cont $\ge 0.898$) by more than 1$\sigma$. Thus 2M~1726+15 is assigned an $H$-band continuum gravity score of 1.}
\end{itemize}

The median gravity score from the 4 indicators above (12?1) is 1, thus we assign the low-resolution spectrum of 2M~1726+15 a gravity classification of {\sc int-g}.

\subsection{A Moderate-resolution Spectrum}

To illustrate a gravity classification for a moderate-resolution spectrum, we examine the spectrum of 2MASS~J10224821+5825453 (L1).  Tables \ref{tbl:sxdews} and \ref{tbl:sxd} list the EWs and indices calculated from this spectrum.  The gravity scores for VO and $H$-band continuum shape are determined in the same way as for the low-resolution spectrum example.  The FeH score will include the FeH$_z$ and FeH$_J$ indices.  The alkali line score will be determined from the EWs of the \ion{Na}{1} and \ion{K}{1} lines, using the criteria in Table \ref{tbl:ewbounds}.

\begin{itemize}
\item{FeH:  the FeH gravity score for moderate-resolution spectra is determined from both the FeH$_z$ and FeH$_J$ indices.  2M~1022+58 has an FeH$_z$ index ($1.284\pm0.031$) that is consistent with normal field L1 dwarfs and receives a score of 0.  Its FeH$_J$ index ($1.229\pm0.024$), however, meets the criteria to be scored a 1 (FeH$_J \le1.253$) by more than 1$\sigma$.  Because one of the FeH indices receives a score of 1, the gravity score for FeH is 1.}
\item{VO:  the VO gravity score is determined from the VO$_z$ index.  The VO$_z$ index of 2M~1022+58 ($1.183\pm0.020$) meets the criteria to be assigned a score of 1 (VO$_z \ge 1.112$) by more than 1$\sigma$.  Thus, the VO gravity score is 1.}
\item{Alkali Lines:  the alkali line gravity score for moderate-resolution spectra is determined from the EWs of the $J$-band \ion{K}{1} and \ion{Na}{1} lines.   Comparing the EWs calculated from the spectrum of 2M1022+58 (Table~\ref{tbl:sxdews}) to the criteria in Table \ref{tbl:ewbounds}, the source receives scores of 1, 0, 0, and 0 for the 1.138~$\mu$m \ion{Na}{1}, 1.169~$\mu$m \ion{K}{1}, 1.177~$\mu$m \ion{K}{1}, and 1.253~$\mu$m \ion{K}{1} line EWs, respectively.  Because it did not receive a score of 1 from at least half of its line EWs, it receives an alkali line gravity score of 0.}
\item{$H$-band continuum shape:  the gravity score for $H$-band continuum shape is determined from the $H$-cont index.  The $H$-cont index value of 2M~1022+58 ($0.913\pm0.008$) does not meet the criteria for an L1 to receive a score of 1, and is thus assigned an $H$-band continuum gravity score of 0.}
\end{itemize}

The median gravity score from the four indicators above (1100) is 0.5.  Thus, the moderate-resolution spectrum of 2M~1022+58 is assigned a gravity classification of {\sc fld-g}.

\clearpage

\begin{deluxetable}{llllllllllll}
\tablecaption{Near-IR Spectral Types}
\tabletypesize{\scriptsize}
\rotate
\tablewidth{0pt}
\tablehead{
  \colhead{Object} &
  \colhead{SpT} &
  \multicolumn{3}{c}{References} &
  \multicolumn{2}{c}{SpT (Visual)} &
  \multicolumn{4}{c}{SpT (Index)} &
  \colhead{SpT} \\
  \colhead{} &
  \colhead{Optical} &
  \colhead{Opt. SpT} &
  \colhead{Youth} &
  \colhead{NIR Spec.} &
  \colhead{$J$-band\tablenotemark{a}} &
  \colhead{$K$-band\tablenotemark{a}} &
  \colhead{H$_2$O} &
  \colhead{H$_2$OD} &
  \colhead{H$_2$O-1} &
  \colhead{H$_2$O-2} &
  \colhead{Final}
}
\startdata
\multicolumn{12}{c}{Low resolution near-IR spectra} \\
\tableline
   2MASS J00274197+0503417\tablenotemark{b}&         M9.5 &     K95 & K95;M99 &     A13 &    L0.0 &    L0.0 & M9.0$\pm$0.6 & M7.0$\pm$1.0 & L0.5$\pm$1.1 & M9.3$\pm$0.7 & L0\\
   2MASS J00325584-4405058 & L0.0$\gamma$ & R08;C09 &     C09 &     A13 &    L1.5 &    L0.0 & M9.9$\pm$0.6 & M9.8$\pm$0.8 & L0.0$\pm$1.1 & M9.3$\pm$0.6 & L0\\
   2MASS J00332386-1521309 &  L4.0$\beta$ & R08;C09 &     C09 &     A13 &    L1.5 &    L1.0 & L0.4$\pm$0.5 & L1.9$\pm$0.8 & M8.6$\pm$1.1 & L0.6$\pm$0.6 & L1\\
 WISEP J004701.06+680352.1 &      \nodata & \nodata &     G12 &     G12 &   L8.0: &   L6.0: &      \nodata & L6.2$\pm$0.8 &      \nodata &      \nodata & L7\\
   2MASS J01033203+1935361 &  L6.0$\beta$ & K00;F12 &     F12 &     C04 &    L7.0 &    L6.0 &      \nodata & L5.2$\pm$0.8 &      \nodata &      \nodata & L6\\
   2MASS J01174748-3403258 &         L2.0 &     C03 &     F09 &     B08 &    L3.0 &    L2.0 & L0.8$\pm$0.6 & L1.5$\pm$0.8 & M9.9$\pm$1.1 & L1.6$\pm$0.6 & L1\\
   2MASS J01262109+1428057 & L4.0$\gamma$ &     F13 &     M08 &     M08 &   L6.0: &    L1.5 & L2.7$\pm$1.2 & L2.5$\pm$0.9 & L0.1$\pm$1.3 & M8.8$\pm$0.8 & L2\\
\enddata

\tablenotetext{a}{Uncertainties in the spectral types are $\pm$1 subtype except where noted by a colon, which indicates an uncertainty of $\pm$2 subtypes.}
\tablenotetext{b}{More commonly known as PC~0025+04}

\tablerefs{
A07:\citet{allers07};
A09:\citet{allers09};
A10:\citet{allers10};
A13:This Work;
B06:\citet{burgasser06};
B08:\citet{burgasser08};
B10:\citet{burgasser10};
C03:\citet{cruz03};
C04:\citet{cruz04};
C06:\citet{chiu06};
C07:\citet{cruz07};
C09:\citet{cruz09};
Ch04:\citet{chauvin04};
Ch05a:\citet{chauvin05a};
Ch05b:\citet{chauvin05b};
F09:\citet{faherty09};
F12:\citet{faherty12};
F13:\citet{faherty13};
G02:\citet{gizis02};
G11:\citet{geissler11};
G12:\citet{gizis12};
H02:\citet{hawley02};
K95:\citet{kirkpatrick95};
K00:\citet{kirkpatrick00};
K01:\citet{kirkpatrick01};
K06:\citet{kirkpatrick06};
K08:\citet{kirkpatrick08};
K10:\citet{kirkpatrick10};
Ka08:\citet{kastner08};
L07:\citet{looper07};
L08:\citet{looper08};
Lu07:(K.~L.~Luhman 2007, private communication);
M99a:\citet{martin99a};
M99b:\citet{martin99};
M03:\citet{mclean03};
M08:\citet{metchev08};
M10:\citet{martin10};
P12:\citet{patience12};
R98:\citet{rebolo98};
R02:\citet{reid02};
R08:\citet{reid08};
R09:\citet{reiners09};
R10:\citet{rice10};
S03:\citet{salim03};
S05:\citet{scholz05};
S07:\citet{schmidt07};
S09:\citet{shkolnik09};
S10:\citet{seifahrt10};
T03:\citet{torres03};
T97:\citet{thackrah97};
T98:\citet{tinney98};
W99:\citet{webb99};
W08:\citet{west08};
W11:\citet{wahhaj11};
W13:\citet{weinberger13}
}

\label{tbl:spt}
\end{deluxetable}

\begin{deluxetable}{llccccccc}
\tablecaption{Log of SpeX Observations}
\tabletypesize{\footnotesize}
\tablewidth{0pt}
\tablehead{
  \colhead{Object} &
  \colhead{Date} &
  \colhead{Mode} &
  \colhead{Slit Width} &
  \colhead{Exposure Time} &
  \colhead{A0 V Standard} \\
  \colhead{} &
  \colhead{(UT)} &
  \colhead{} &
  \colhead{(arcsec)} &
  \colhead{(s)} &
  \colhead{} \\
}
\startdata
2MASS J00034227-2822410  & 2006 Sep 13 & SXD & 0.5 & 720 & HD 141 \\ 
2MASS J00192626+4614078  & 2010 Sep 10 & SXD & 0.5 &  480 & HD 1561 \\ 
2MASS J00274197+0503417  & 2012 Sep 20 & PRZ & 0.5 & 2160  & HD 1663 \\ 
2MASS J00325584-4405058  & 2008 Aug 15 & PRZ & 0.5 & 1800 & HD 8977\\ 
2MASS J00332386-1521309   & 2008 Nov 29 & PRZ & 0.5 & 900  & HD 222332 \\ 
2MASS J00452143+1634446  & 2008 Aug 15 & SXD & 0.8 & 1440 & HD 7215\\ 
SERC 296A                              & 2006 Nov 19 & SXD & 0.5 & 1200 & HD18546 \\ 
2MASS J02292794-0053282  &  2012 Sep 20 & PRZ & 0.5 & 1080 & HD 18571 \\
\enddata

\label{tbl:spex}
\end{deluxetable}

\begin{deluxetable}{llllllll}
\tablecaption{Spectral Type Sensitive Indices}
\tabletypesize{\footnotesize}
\tablewidth{0pt}
\tablehead{
  \colhead{Index Definition} &
  \colhead{Index Ref.} &
  \colhead{SpT} &
  \multicolumn{4}{c}{Coefficients of Polynomial Fits\tablenotemark{a}} &
  \colhead{rms} \\
  \colhead{} &
  \colhead{} &
  \colhead{}  &
  \colhead{$c_0$} &
  \colhead{$c_1$} &
  \colhead{$c_2$} &
  \colhead{$c_3$} &
  \colhead{SpT} \\
}
\startdata
${\rm H_2O} = \frac{\langle F_{\lambda=1.550--1.560} \rangle}{\langle F_{\lambda=1.492--1.502} \rangle}$     & \citet{allers07} & M5--L4 &-83.5437 & 169.388 & -104.424 & 24.0476 & 0.390 \\
${\rm H_2OD} = \frac{\langle F_{\lambda=1.951--1.977} \rangle}{\langle F_{\lambda=2.062--2.088} \rangle}$& \citet{mclean03} & L0--L8 & 79.4477 & -202.245 & 229.884 & -97.230 & 0.757 \\
${\rm H_2O}$-1$ = \frac{\langle F_{\lambda=1.335--1.345} \rangle}{\langle F_{\lambda=1.295--1.305} \rangle}$& \citet{slesnick04}& M4--L5 & 12.1927 & 39.3513 & -80.7404 & 28.5982 & 1.097 \\
${\rm H_2O}$-2$ = \frac{\langle F_{\lambda=2.035--2.045} \rangle}{\langle F_{\lambda=2.145--2.155} \rangle}$ & \citet{slesnick04}& M4--L2 & 10.8822 & 55.4580 &-97.8144 & 37.5013 & 0.501 \\
\enddata

\tablenotetext{a}{The spectral type is calculated from the polynomial fits as: SpT~$= c_0 + c_1 ({\rm index}) + c_2 ({\rm index})^2 + c_3 ({\rm index})^3$.}

\label{tbl:indexfits}
\end{deluxetable}

\begin{deluxetable}{lrrrr}
\tablecaption{Gravity-sensitive Spectral Indices}
\tabletypesize{\footnotesize}
\tablewidth{0pt}
\tablehead{
  \colhead{Index\tablenotemark{a}} &
  \colhead{$\lambda_{\rm{line}}$} &
  \colhead{$\lambda_{\rm{cont1}}$} &
  \colhead{$\lambda_{\rm{cont2}}$} &
  \colhead{Bandwidth}\\
  \colhead{} &
  \colhead{($\mu$m)} &
  \colhead{($\mu$m)} &
  \colhead{($\mu$m)} &
  \colhead{($\mu$m)} \\
} 
\startdata
FeH$_z$ & 0.998 & 0.980 & 1.022 & 0.0133 \\
VO$_z$  & 1.058 & 1.035 & 1.087 & 0.0141 \\
FeH$_J$ & 1.200 & 1.192 & 1.208 & 0.0024 \\
\ionfootnote{K}{I}$_J$& 1.244 & 1.220 & 1.270 & 0.0166 \\
$H$--cont & 1.560 & 1.470 & 1.670 & 0.0208 \\
\enddata

\tablenotetext{a}{The method for calculating indices is described in Section 4.2.1.} 
\label{tbl:indices}
\end{deluxetable}

\begin{deluxetable}{lllrrrrcc}
\tablecaption{Gravity Classifications for Low-resolution Spectra}
\tabletypesize{\scriptsize}
\rotate
\tablewidth{0pt}
\tablehead{
  \colhead{Object} &
  \multicolumn{2}{c}{SpT} &
  \multicolumn{4}{c}{Index Values} &
  \colhead{Gravity} &
  \colhead{Gravity} \\
  \colhead{} &
  \colhead{Optical} &
  \colhead{Near-IR} &
  \colhead{FeH$_z$} &
  \colhead{VO$_z$} &
  \colhead{\ionscript{K}{I}$_J$} &
  \colhead{$H$--cont} &
  \colhead{Scores\tablenotemark{a}} &
  \colhead{Class} \\
}
\startdata
   2MASS J00274197+0503417 &         M9.5 & L0 & $ 1.127\pm 0.019$ & $ 1.186\pm 0.025$ & $ 1.071\pm 0.015$ & $ 1.010\pm 0.014$ & 1112 & {\sc   int-g}\\
   2MASS J00325584-4405058 & L0.0$\gamma$ & L0 & $ 1.098\pm 0.028$ & $ 1.233\pm 0.016$ & $ 1.077\pm 0.012$ & $ 0.973\pm 0.008$ & 2112 & {\sc    vl-g}\\
   2MASS J00332386-1521309 &  L4.0$\beta$ & L1 & $ 1.300\pm 0.066$ & $ 1.047\pm 0.019$ & $ 1.126\pm 0.024$ & $ 0.916\pm 0.010$ & 000? & {\sc   fld-g}\\
 WISEP J004701.06+680352.1 &      \nodata & L7 & $ 1.060\pm 0.024$ & $ 0.991\pm 0.023$ & $ 1.026\pm 0.010$ & $ 0.912\pm 0.014$ & nnn1 & {\sc \nodata}\\
   2MASS J01033203+1935361 &  L6.0$\beta$ & L6 & $ 1.038\pm 0.031$ & $ 0.971\pm 0.020$ & $ 0.987\pm 0.024$ & $ 0.905\pm 0.018$ & nnn1 & {\sc \nodata}\\
   2MASS J01174748-3403258 &         L2.0 & L1 & $ 1.161\pm 0.025$ & $ 1.243\pm 0.051$ & $ 1.062\pm 0.018$ & $ 0.962\pm 0.013$ & 1121 & {\sc   int-g}\\
   2MASS J01262109+1428057 & L4.0$\gamma$ & L2 & $ 1.112\pm 0.079$ & $ 1.085\pm 0.041$ & $ 1.036\pm 0.049$ & $ 0.966\pm 0.028$ & 2022 & {\sc    vl-g}\\
   2MASS J01415823-4633574 & L0.0$\gamma$ & L0 & $ 1.066\pm 0.021$ & $ 1.371\pm 0.038$ & $ 1.044\pm 0.022$ & $ 1.003\pm 0.008$ & 2222 & {\sc    vl-g}\\
\enddata

\tablenotetext{a}{Gravity scores are listed in the following order:  FeH, VO, alkali lines, and $H$-band continuum shape. See Section 4.3 for details.}

\label{tbl:prz}
\end{deluxetable}

\begin{deluxetable}{lllrrrrcc}
\tablecaption{Gravity Classifications for Moderate-resolution Spectra}
\tabletypesize{\scriptsize}
\rotate
\tablewidth{0pt}
\tablehead{
  \colhead{Object} &
  \multicolumn{2}{c}{SpT} &
  \multicolumn{4}{c}{Index Values} &
  \colhead{Gravity} &
  \colhead{Gravity} \\
  \colhead{} &
  \colhead{Optical} &
  \colhead{Near-IR} &
  \colhead{FeH$_z$} &
  \colhead{FeH$_J$} &
  \colhead{VO$_z$} &
  \colhead{$H$--cont} &
  \colhead{Scores\tablenotemark{a}} &
  \colhead{Class} \\
}
\startdata
   2MASS J00034227-2822410 &         M7.5 & M7 &$ 1.125\pm 0.015$ &$ 1.100\pm0.012$ &$ 1.035\pm 0.010$ &$ 0.966\pm 0.007$ &0n00 &{\sc   fld-g}\\
   2MASS J00192626+4614078 &         M8.0 & M8 &$ 1.120\pm 0.012$ &$ 1.113\pm0.011$ &$ 1.040\pm 0.008$ &$ 0.973\pm 0.005$ &1n11 &{\sc   int-g}\\
   2MASS J00452143+1634446 &  L2.0$\beta$ & L2 &$ 1.187\pm 0.015$ &$ 1.192\pm0.016$ &$ 1.283\pm 0.014$ &$ 0.942\pm 0.007$ &1221 &{\sc    vl-g}\\
   2MASS J01033203+1935361 &  L6.0$\beta$ & L6 &$ 1.156\pm 0.011$ &$ 1.098\pm0.011$ &$ 1.015\pm 0.010$ &$ 0.896\pm 0.007$ &1n11 &{\sc   int-g}\\
                 SERC 296A &         M6.0 & M6 &$ 1.065\pm 0.006$ &$ 1.062\pm0.006$ &$ 1.010\pm 0.005$ &$ 0.996\pm 0.005$ &0n22 &{\sc    vl-g}\\
   2MASS J01415823-4633574 & L0.0$\gamma$ & L0 &$ 1.085\pm 0.022$ &$ 1.122\pm0.018$ &$ 1.383\pm 0.024$ &$ 1.001\pm 0.009$ &2212 &{\sc    vl-g}\\
          GSC 08047-00232B &      \nodata & L1 &          \nodata &$ 1.120\pm0.055$ &          \nodata &$ 0.953\pm 0.010$ &2n21 &{\sc    vl-g}\\
    2MASSI J0253597+320637 &         M7.5 & M6 &$ 1.083\pm 0.009$ &$ 1.060\pm0.012$ &$ 1.046\pm 0.007$ &$ 0.972\pm 0.004$ &0n00 &{\sc   fld-g}\\
\enddata
\tablenotetext{a}{Gravity scores are listed in the following order:  FeH, VO, alkali lines, and $H$-band continuum shape. See Section 4.3 for details.}
\label{tbl:sxd}
\end{deluxetable}


\begin{deluxetable}{crrrrrr}
\tablecaption{Line Equivalent Width Calculation}
\tabletypesize{\footnotesize}
\tablewidth{0pt}
\tablehead{
  \colhead{Species} &
  \colhead{$\lambda_{\rm{line}}$} &
  \colhead{$\lambda_{\rm{cont1}}$} &
  \colhead{$\lambda_{\rm{cont2}}$} &
  \colhead{Window$_{\rm{line}}$} &
  \colhead{Window$_{\rm{cont}}$} \\
  \colhead{} &
  \colhead{($\mu$m)} &
  \colhead{($\mu$m)} &
  \colhead{($\mu$m)} &
  \colhead{($\mu$m)} &
  \colhead{($\mu$m)} \\
} 
\startdata
\ionfootnote{Na}{I} & 1.1396 & 1.1310 & 1.1470 & 0.006 & 0.002\\
\ionfootnote{K}{I} & 1.1692 & 1.1630 & 1.1860 & 0.006 & 0.002\\
\ionfootnote{K}{I} & 1.1778 & 1.1630 & 1.1860 & 0.006 & 0.002\\
\ionfootnote{K}{I} & 1.2437 & 1.2480 & 1.2570 & 0.006 & 0.002\\
\ionfootnote{K}{I} &  1.2529 & 1.2480 & 1.2570 & 0.006 & 0.002\\
\enddata

\tablenotetext{a}{The method for calculating equivalent widths is described in Section 4.2.3.} 
\label{tbl:ews}
\end{deluxetable}

\begin{deluxetable}{lllrrrr}
\tablecaption{Equivalent Widths Calculated from Moderate-resolution Spectra}
\tabletypesize{\scriptsize}
\tablewidth{0pt}
\tablehead{
  \multicolumn{2}{c}{SpT} &
  \colhead{Near-IR} &
  \colhead{\ionscript{Na}{I} 1.138~$\mu$m} &
  \colhead{\ionscript{K}{I} 1.169~$\mu$m} &
  \colhead{\ionscript{K}{I} 1.177~$\mu$m} &
  \colhead{\ionscript{K}{I} 1.253~$\mu$m} \\
  \colhead{} &
  \colhead{Optical} & 
  \colhead{Near-IR} &
  \colhead{EW (\AA)} &
  \colhead{EW (\AA)} &
  \colhead{EW (\AA)} &
  \colhead{EW (\AA)} \\
}
\startdata
   2MASS J00034227-2822410 &         M7.5 & M7 & $ 13.02\pm 0.32$ & $  3.06\pm 0.63$ & $  6.02\pm 0.63$ & $  4.65\pm 0.50$ \\
   2MASS J00192626+4614078 &         M8.0 & M8 & $ 10.20\pm 0.33$ & $  3.44\pm 0.42$ & $  4.87\pm 0.44$ & $  3.65\pm 0.34$ \\
   2MASS J00452143+1634446 &  L2.0$\beta$ & L2 & $  7.63\pm 0.45$ & $  4.84\pm 0.46$ & $  5.86\pm 0.46$ & $  3.90\pm 0.38$ \\
   2MASS J01033203+1935361 &  L6.0$\beta$ & L6 & $  5.28\pm 1.91$ & $  6.39\pm 0.56$ & $  6.50\pm 0.53$ & $  5.75\pm 0.27$ \\
                 SERC 296A &         M6.0 & M6 & $  6.22\pm 0.39$ & $  1.37\pm 0.26$ & $  2.49\pm 0.27$ & $  1.63\pm 0.21$ \\
   2MASS J01415823-4633574 & L0.0$\gamma$ & L0 & $  7.70\pm 1.22$ & $  3.77\pm 1.02$ & $  5.28\pm 1.05$ & $  1.92\pm 0.66$ \\
          GSC 08047-00232B &      \nodata & L1 &          \nodata & $  1.75\pm 2.16$ & $  2.63\pm 2.17$ & $  1.76\pm 1.32$ \\
    2MASSI J0253597+320637 &         M7.5 & M6 & $ 12.66\pm 0.33$ & $  3.58\pm 0.31$ & $  5.32\pm 0.29$ & $  2.78\pm 0.49$ \\
\enddata
\label{tbl:sxdews}
\end{deluxetable}

\begin{deluxetable}{lrrrrrrrrrr}
\tablecaption{Determination of Gravity Scores from Indices}
\tabletypesize{\scriptsize}
\tablewidth{0pt}
\tablehead{
  \colhead{SpT} &
  \multicolumn{2}{c}{FeH$_z$} &
  \multicolumn{2}{c}{FeH$_J$} &
  \multicolumn{2}{c}{VO$_z$} &
  \multicolumn{2}{c}{\ionscript{K}{I}$_J$} &
  \multicolumn{2}{c}{$H$-cont} \\
  \colhead{Near-IR} &
  \colhead{1} &
  \colhead{2} &
  \colhead{1} &
  \colhead{2} &
  \colhead{1} &
  \colhead{2} &
  \colhead{1} &
  \colhead{2} &
  \colhead{1} &
  \colhead{2} \\
}
\startdata
M5 &     \nodata &     \nodata &     \nodata &     \nodata &     \nodata &     \nodata &     \nodata &     \nodata &     \nodata &     \nodata \\
M6 &  $\le$1.068 &  $\le$1.039 &  $\le$1.049 &  $\le$1.040 &     \nodata &     \nodata &  $\le$1.042 &  $\le$1.028 &  $\ge$0.988 &  $\ge$0.994 \\
M7 &  $\le$1.103 &  $\le$1.056 &  $\le$1.079 &  $\le$1.056 &     \nodata &     \nodata &  $\le$1.059 &  $\le$1.036 &  $\ge$0.981 &  $\ge$0.990 \\
M8 &  $\le$1.146 &  $\le$1.074 &  $\le$1.116 &  $\le$1.076 &     \nodata &     \nodata &  $\le$1.077 &  $\le$1.046 &  $\ge$0.963 &  $\ge$0.984 \\
M9 &  $\le$1.167 &  $\le$1.086 &  $\le$1.160 &  $\le$1.100 &     \nodata &     \nodata &  $\le$1.085 &  $\le$1.053 &  $\ge$0.949 &  $\ge$0.979 \\
L0 &  $\le$1.204 &  $\le$1.106 &  $\le$1.208 &  $\le$1.124 &  $\ge$1.122 &  $\ge$1.256 &  $\le$1.098 &  $\le$1.061 &  $\ge$0.935 &  $\ge$0.972 \\
L1 &  $\le$1.252 &  $\le$1.121 &  $\le$1.253 &  $\le$1.144 &  $\ge$1.112 &  $\ge$1.251 &  $\le$1.114 &  $\le$1.067 &  $\ge$0.914 &  $\ge$0.968 \\
L2 &  $\le$1.298 &  $\le$1.142 &  $\le$1.288 &  $\le$1.158 &  $\ge$1.110 &  $\ge$1.232 &  $\le$1.133 &  $\le$1.073 &  $\ge$0.906 &  $\ge$0.964 \\
\enddata
\label{tbl:indexbounds}
\end{deluxetable}

\begin{deluxetable}{lrrrrrrrr}
\tablecaption{Determination of Gravity Scores from Equivalent Widths}
\tabletypesize{\scriptsize}
\tablewidth{0pt}
\tablehead{
  \colhead{SpT} &
  \multicolumn{2}{c}{\ionscript{Na}{I} 1.138~$\mu$m} &
  \multicolumn{2}{c}{\ionscript{K}{I} 1.169~$\mu$m} &
  \multicolumn{2}{c}{\ionscript{K}{I} 1.177~$\mu$m} &
  \multicolumn{2}{c}{\ionscript{K}{I} 1.253~$\mu$m} \\
  \colhead{Near-IR} &
  \multicolumn{2}{c}{EW (\AA)} &
  \multicolumn{2}{c}{EW (\AA)} &
  \multicolumn{2}{c}{EW (\AA)} &
  \multicolumn{2}{c}{EW (\AA)} \\
  \colhead{} &
  \colhead{1} &
  \colhead{2} &
  \colhead{1} &
  \colhead{2} &
  \colhead{1} &
 \colhead{2} &
  \colhead{1} &
  \colhead{2} \\
}
\startdata
M5 & $\le$ 6.128 & $\le$ 4.402 &     \nodata &     \nodata &     \nodata &     \nodata & $\le$ 0.905 & $\le$ 1.112\\
M6 & $\le$ 9.087 & $\le$ 5.509 & $\le$ 1.632 & $\le$ 1.524 & $\le$ 3.124 & $\le$ 2.521 & $\le$ 2.215 & $\le$ 1.696\\
M7 & $\le$10.742 & $\le$ 6.132 & $\le$ 2.816 & $\le$ 2.102 & $\le$ 4.690 & $\le$ 3.255 & $\le$ 3.277 & $\le$ 2.190\\
M8 & $\le$11.700 & $\le$ 6.511 & $\le$ 3.954 & $\le$ 2.665 & $\le$ 6.067 & $\le$ 3.903 & $\le$ 4.214 & $\le$ 2.642\\
M9 & $\le$12.318 & $\le$ 6.770 & $\le$ 5.074 & $\le$ 3.222 & $\le$ 7.290 & $\le$ 4.483 & $\le$ 5.094 & $\le$ 3.072\\
L0 & $\le$12.754 & $\le$ 6.948 & $\le$ 6.179 & $\le$ 3.767 & $\le$ 8.396 & $\le$ 5.012 & $\le$ 5.937 & $\le$ 3.483\\
L1 & $\le$13.018 & $\le$ 7.022 & $\le$ 7.244 & $\le$ 4.282 & $\le$ 9.408 & $\le$ 5.495 & $\le$ 6.727 & $\le$ 3.856\\
L2 & $\le$12.629 & $\le$ 6.934 & $\le$ 8.223 & $\le$ 4.736 & $\le$10.325 & $\le$ 5.924 & $\le$ 7.414 & $\le$ 4.161\\
\enddata
\label{tbl:ewbounds}
\end{deluxetable}

\begin{deluxetable}{lcccll}
\tablecaption{Young Field Objects with Age Estimates}
\tabletypesize{\scriptsize}
\tablewidth{0pt}
\tablecolumns{2}
\tablehead{
  \colhead{Object} &
  \colhead{SpT} &
  \colhead{Gravity} &
  \colhead{Age} &
  \colhead{Notes} &
  \colhead{Ref.} \\
  \colhead{} &
  \colhead{Near-IR} &
  \colhead{Near-IR} &
  \colhead{(Myr)} &
 \colhead{} &
  \colhead{} \\
} 
\startdata
TWA 28 & M9 & {\sc vl-g} & $\sim$10 & TWA member &  S05\\
TWA 5B & M9 & {\sc vl-g} & $\sim$10 & TWA member & W99\\
TWA 8B & M6 & {\sc vl-g} & $\sim$10 &TWA member & W99 \\
TWA 26 & M9 & {\sc vl-g} & $\sim$10 &TWA member & G02 \\
TWA 27A & M8 & {\sc vl-g} & $\sim$10 &TWA member & G02 \\
TWA 27B & L3 & {\sc vl-g} & $\sim$10 &TWA member & Ch04 \\
TWA 11C & M5 & {\sc vl-g} & $\sim$10 & TWA member & Ka08;W13\\
TWA 29 & L0 & {\sc vl-g} & $\sim$10 &TWA member & L07;W13 \\
\enddata
\label{tbl:ages}
\tablerefs{
A10:\citet{allers10};
Ch04:\citet{chauvin04};
Ch05a:\citet{chauvin05a};
Ch05b:\citet{chauvin05b};
G02:\citet{gizis02};
K01:\citet{kirkpatrick01};
K06:\citet{kirkpatrick06};
Ka08:\citet{kastner08};
L13:\citet{liu13};
L07:\citet{looper07};
R98:\citet{rebolo98};
R09:\citet{reiners09};
R10:\citet{rice10};
S03:\citet{salim03};
S05:\citet{scholz05};
S09:\citet{shkolnik09};
S10:\citet{seifahrt10};
S12:\citet{shkolnik12};
T97:\citet{thackrah97};
W99:\citet{webb99};
W11:\citet{wahhaj11};
W13:\citet{weinberger13}.
}

\end{deluxetable}

\clearpage

\clearpage

\clearpage

\end{document}